\documentclass{article}

\usepackage{preamble} 

\title{A Regularised Latent-Class Item Response Model for Detecting Measurement Non-Invariance in Ordinal Response Scales}

\author[1]{Gabriel Wallin}
\author[2]{ Qi Huang}

\affil[1]{School of Mathematical Sciences, Lancaster University}
\affil[2]{Department of Educational Studies, Purdue University}
\date{}

\begin{document}

\maketitle
		
\begin{abstract}
Measurement non-invariance arises when the psychometric properties of a scale differ across subgroups, undermining the validity of group comparisons. At the item level, this manifests as differential item functioning (DIF), where item responses differ across groups after controlling for the latent trait. This paper develops a framework for detecting DIF in ordinal scales without requiring known group labels or anchor items. We formulate a proportional-odds latent-class item response model in which individuals are assigned probabilistically to latent classes. DIF is captured through class-specific intercept and slope shifts, allowing both uniform and non-uniform DIF. Identification is achieved through an \(\ell_1\)-penalised marginal likelihood under a sparsity assumption, with estimation implemented using a tailored EM algorithm. Because class-specific slopes leave both the location and scale of each latent class unidentified, sparsity anchors the latent metric while selecting DIF effects. Simulation studies demonstrate accurate recovery of item parameters and both types of DIF. An empirical application to a personality test reveals latent subgroups with distinct response patterns and identifies items displaying potential class-specific measurement non-invariance. The framework provides a flexible approach for assessing measurement invariance in ordinal scales when comparison groups are unobserved or poorly defined.
\end{abstract}



\section{Introduction}\label{sec:introduction}

Psychometric instruments such as personality tests, attitude scales, and questionnaires are widely used to infer latent variables across the social and behavioural sciences \citep{embretson2013item}. In many situations, respondent populations are heterogeneous: individuals may differ in culture, language, age, gender, or educational background. Meaningful comparisons across such groups require that the instrument measures the same construct in the same way for all respondents. When this property, known as measurement invariance, is violated, group comparisons may reflect measurement artifacts rather than purely true differences in the underlying latent trait \citep{millsap2012statistical, steenkamp1998assessing}.

At the item level, measurement non-invariance manifests as differential item functioning (DIF): respondents with equal latent trait levels but from different subgroups have systematically different probabilities of endorsing particular items \citep{millsap2012statistical}. 
Methods for detecting DIF have a long history in educational measurement and psychometrics \citep[e.g.,][]{lord1980applications, holland1986differential, holland1993differential}. Broadly, DIF detection procedures can be categorised into item response theory (IRT)-based methods and non-IRT approaches. In this paper, we focus on an IRT-based framework, which utilizes a probabilistic link between latent traits and observed responses \citep{birnbaum1968some, embretson2013item, samejima1969estimation}. 

Classical DIF procedures assume that respondent groups are observed and that a set of DIF-free anchor items is known in advance \citep{millsap2012statistical}. Anchor items identify the latent trait scale, and group labels enable estimation of group-specific parameters. However, these requirements are often unrealistic in modern applications where relevant grouping structures may be unknown, partially observed, intersectional, or multidimensional. Even when demographic information is available, it is rarely obvious which variable, or combination of variables, meaningfully defines the comparison groups driving DIF. Likewise, anchor sets are difficult to specify a priori, and misspecifying them can bias parameter estimates and inflate false discoveries \citep{kopf2015anchor, kopf2015framework, wang2003effects}. {The motivation for considering latent comparison groups is not that such groups are guaranteed to correspond to substantively labelled populations. Rather, the aim is to detect otherwise hidden forms of measurement heterogeneity that may not align with any single observed covariate. In many applications, response behaviour may be shaped by combinations of demographic characteristics, response styles, cultural background, language use, or unmeasured contextual factors. A latent-class DIF model therefore provides an exploratory device for identifying candidate patterns of non-invariance when the relevant grouping structure is unknown. The resulting classes should be interpreted as model-based response groups whose substantive meaning requires external validation, rather than being assumed by the model. From a measurement perspective, however, their value does not depend on whether they can be given clear substantive labels. What matters is that leaving such heterogeneity unmodeled may bias the estimated trait distribution and any comparisons drawn from it.}

{Building on the broader mixture and heterogeneous-population IRT literature, including early work on latent variable models for heterogeneous populations and measurement invariance \citep{muthen1989latent}}, two main strands of DIF research have relaxed one of these requirements while retaining the other. Latent DIF approaches allow group membership to be unobserved, typically through latent class or mixture IRT models, but require a fixed anchor set \citep{cohen2005mixture, de2011explanatory, cho2010multilevel, cho2016ncme}. These models are particularly relevant when demographic variables do not align with meaningful measurement subgroups; for instance, when personality response patterns reflect latent temperament subtypes or when cultural heterogeneity creates measurement differences that do not map onto observed nationality or language \citep{teresi2016epilogue, reeve2016overview, teresi2021differential}. By contrast, regularisation-based approaches treat anchor identification as a variable selection problem and impose sparsity on DIF parameters so that most items are invariant \citep{tibshirani1996regression, magis2015detection, tutz2015penalty, belzak2020improving}. Extensions have addressed non-uniform DIF \citep{woods2011testing}, MIMIC frameworks \citep{woods2009evaluation, wang2009mimic}, and inference without known anchors \citep{chen2023dif, yuan2021differential}. However, most regularised DIF procedures require observed groups, focus primarily on binary items, and often target only uniform DIF.

Recent work has expanded the scope of DIF detection toward complex population structures and multigroup settings. For multi-study data harmonisation, \citet{lyu2025multi} developed a regularised variational EM framework for multidimensional IRT to detect DIF across multiple studies efficiently. \citet{lyu2025detecting} proposed a group pairwise penalty for 2PL models that avoids shrinkage toward a single reference group, improving sensitivity when multiple small comparison groups are present. \citet{bauer2020simplifying} introduced regularised moderated nonlinear factor analysis to assess measurement invariance across multiple background variables simultaneously. \citet{ren2025novel} introduced a multilevel random item effects model using regularised Gaussian variational estimation to detect DIF arising from interactions between identity attributes, with extensions to uniform and non-uniform DIF. Collectively, these methods illustrate a growing trend among DIF procedures that move beyond the traditional reference versus focal paradigm and accommodate increasingly rich group structures.

Despite this progress, most existing approaches still presuppose explicit grouping variables or a known set of anchor items. \citet{wallin2024dif}, however, relaxed both classical requirements simultaneously by combining latent class assignment with $\ell_1$-penalised DIF detection for binary items under the two-parameter logistic (2PL) model. By treating both group membership and anchor status as unknown, the method identifies latent subgroups and DIF items jointly through regularised marginal maximum likelihood estimation, and computation was carried out using an expectation maximisation \citep[EM;][]{dempster1977maximum, bock1981marginal} algorithm. However, this framework addresses only binary responses and uniform DIF (intercept shifts), leaving ordinal data and non-uniform DIF (slope differences) outside its scope. {The present extension is substantive for several reasons. 
Moving from binary responses to ordinal responses is not only a change in outcome type: ordinal instruments require modelling ordered response categories through item-specific threshold parameters, and dichotomising such responses may discard information about response intensity and category structure. More consequentially, allowing non-uniform DIF changes the identification problem itself, not merely its dimensionality. In the uniform-DIF-only setting of Wallin et al (2024), the only source of within-class non-identifiability is a location shift in the latent trait distribution, which can be absorbed by the intercept (uniform) DIF parameters. Once class-specific slopes are permitted, a second indeterminacy arises: a scale transformation of the class-specific latent distribution can be absorbed by the slope (non-uniform) DIF parameters. We make this explicit in Section 2.5.1, where we show that a location–scale reparametrisation of any non-reference class can be exactly offset by compensating shifts in both the uniform and non-uniform DIF parameters, leaving the marginal likelihood unchanged. Consequently, the sparsity assumption must now anchor both the location and the scale of every non-reference class simultaneously, which is a harder anchoring problem than in the uniform-only case. The contribution of the present paper therefore lies not in the use of latent classes within IRT per se, nor simply in accommodating ordinal responses, but in resolving this richer location-and-scale anchoring problem through data-driven sparse selection of both class-specific location and slope DIF.}

A methodological gap thus remains. Existing latent DIF methods require known anchors; regularised approaches assume observed groups; intersectional and multigroup methods presuppose explicit grouping structures even when they may be latent; and nearly all approaches focus primarily on binary items. To our knowledge, no general framework simultaneously accommodates unobserved group membership, unspecified anchor items, ordinal response data, and both uniform and non-uniform DIF. This paper proposes a sparse latent-class ordinal IRT framework that fills this gap. Building on \citet{wallin2024dif}, we develop a proportional-odds mixture formulation in which latent group membership, anchor status, and both uniform and non-uniform DIF effects are unknown and estimated jointly. Respondents are probabilistically assigned to latent classes, and item responses follow a proportional odds structure. Class-specific intercept shifts represent uniform DIF, while class-specific slope shifts represent effects analogous to non-uniform DIF under the inferred latent class structure. An $\ell_1$ penalty is applied to all DIF parameters within a marginal likelihood formulation to encourage sparse DIF patterns, enabling data-driven identification of both anchor items and latent groups. Estimation proceeds via a tailored EM algorithm with a proximal update step to handle the non-smooth penalty component \citep{parikh2014proximal}.

The remainder of the article is organised as follows. Section~\ref{sec:statistical-framework} introduces the model and estimation procedure. Section~\ref{sec:Empirical} applies the method to the Fisher Temperament Inventory. Section~\ref{sec:Simulation} presents simulation studies that assess parameter recovery, classification accuracy, and DIF detection performance under a range of conditions. Section~\ref{sec:Discussion} concludes with a discussion of the implications, limitations, and directions for future research.


\section{Proposed Framework}\label{sec:statistical-framework}

\subsection{Notation}

Let $\mathbf{Y} = \{Y_{ij}: i = 1, \ldots, N, j = 1, \ldots, J\}$ denote the observed response matrix, where $Y_{ij} \in \{1, 2, \ldots, M_j\}$ represents the ordinal response of individual $i$ to item $j$, and $M_j$ denotes the number of ordered response categories for item $j$. Each individual is assumed to belong to one of $K+1$ latent groups, indexed by the unobserved class membership variable $\xi_i \in \{0, 1, \ldots, K\}$, where class 0 is designated as the reference group. The latent trait of individual $i$ is denoted by $\theta_i \in \mathbb{R}$, representing the unidimensional construct measured by the instrument.

\subsection{Measurement Model}

To model ordinal responses and conduct DIF analysis, we employ a proportional odds formulation that allows DIF in both location and slope parameters. The proportional odds model is a natural choice for Likert-type ordinal items, as it preserves the ordered nature of response categories and allows flexible estimation of category thresholds \citep{samejima1969estimation}. 

Let $P_{ijm} = P(Y_{ij} \leq m | \theta_i, \xi_i)$ denote the cumulative probability for individual $i$ responding at or below category $m$ on item $j$. The measurement model is specified as:
\begin{equation}\label{eq:measurement_model}
\text{logit}(P_{ijm}) = \tau_{jm} - (a_j + \delta_{2j\xi_i})\theta_i + \delta_{1j\xi_i}
\end{equation}
for $m = 1, \ldots, M_j - 1$, where $\tau_{jm}$ represents the threshold parameter for category $m$ of item $j$, satisfying the ordering constraint $\tau_{j1} < \tau_{j2} < \cdots < \tau_{j,M_j-1}$. The parameter $a_j > 0$ denotes the baseline slope parameter (also known as the discrimination parameter in psychometric applications) for item $j$, which quantifies the strength of the relationship between the latent trait and the item response. The parameters $\delta_{1j\xi_i}$ and $\delta_{2j\xi_i}$ represent class-specific DIF effects, capturing location (uniform DIF) and slope (non-uniform DIF) shifts, respectively. Note that this formulation is equivalent to the graded response model under a sign change of the threshold parameters.

To ensure model identification, we impose the reference-group constraints $\delta_{1j0} = 0$ and $\delta_{2j0} = 0$ for all $j \in \{1, \ldots, J\}$. Under these constraints, the measurement model for the reference group reduces to an alternative parametrisation of the standard proportional odds model:
\begin{equation}\label{eq:reference_model}
\text{logit}(P_{ij0m}) = \tau_{jm} - a_j\theta_i
\end{equation}

The probability of observing response category $m$ is obtained from the cumulative probabilities as
\[
P(Y_{ij} = m \mid \theta_i, \xi_i) =
\begin{cases}
P_{ij1}, & m = 1, \\
P_{ijm} - P_{ij,m-1}, & m = 2, \ldots, M_j-1, \\
1 - P_{ij,M_j-1}, & m = M_j,
\end{cases}
\]
where $P_{ij0} = 0$ by convention.

\subsection{Structural Model}

We model the latent class membership as
\begin{equation}\label{eq:class_dist}
\xi_i \sim \text{Categorical}(\boldsymbol{\nu}),
\end{equation}
where $\boldsymbol{\nu} = (\nu_0, \nu_1, \ldots, \nu_K)^{\top}$ is a vector of class probabilities satisfying $\nu_k \ge 0$ and $\sum_{k=0}^K \nu_k = 1$, with support on the finite set $\{0, 1, \ldots, K\}$. Conditional on the latent class membership, the latent trait follows a class-specific normal distribution:
\begin{equation}\label{eq:trait_dist}
\theta_i | \xi_i = k \sim \mathcal{N}(\mu_k, \sigma_k^2)
\end{equation}
where $\mu_k \in \mathbb{R}$ and $\sigma_k^2 > 0$ represent the class-specific mean and variance parameters, respectively. For identification purposes, we fix $\mu_0 = 0$ and $\sigma_0^2 = 1$ for the reference group.

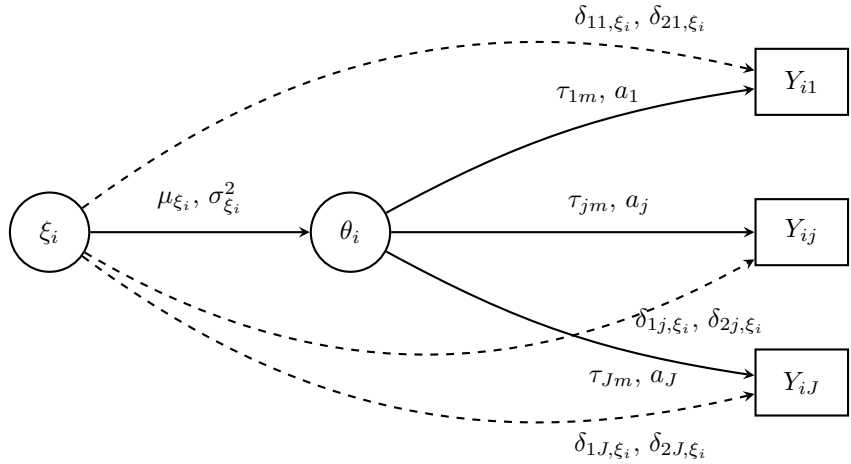
\begin{figure}[ht]
  \centering
\begin{tikzpicture}[
    latent/.style={circle,draw,thick,minimum size=30pt,inner sep=3pt},
    obs/.style={rectangle,draw,thick,minimum width=35pt,minimum height=25pt,inner sep=3pt},
    >=stealth,
    every edge/.style={thick,->,draw=black}
  ]
  \node[latent] (xi)    at (0,0)           {$\xi_i$};
  \node[latent] (theta) at (4,0)           {$\theta_i$};
  
  \node[obs]    (Y1)    at (10,2)   {$Y_{i1}$};
  \node[obs]    (Yj)    at (10,0)   {$Y_{ij}$};
  \node[obs]    (YJ)    at (10,-2)  {$Y_{iJ}$};
  
  \path (xi) edge
        node[above=3pt] {$\mu_{\xi_i},\,\sigma_{\xi_i}^2$}
        (theta);
  
  \path (theta) edge[bend left=10]
        node[pos=0.6, above=3pt] {$\tau_{1m},\,a_1$}
        (Y1);
  \path (theta) edge
        node[pos=0.6, above=3pt] {$\tau_{jm},\,a_j$}
        (Yj);
  \path (theta) edge[bend right=10]
        node[pos=0.7, below=3pt] {$\tau_{Jm},\,a_J$}
        (YJ);
  
  \path (xi) edge[dashed,bend left=25]
        node[pos=0.85, above=3pt] {$\delta_{11,\xi_i},\,\delta_{21,\xi_i}$}
        (Y1);
  \path (xi) edge[dashed,bend right=30]
        node[pos=0.92, below=4pt] {$\delta_{1j,\xi_i},\,\delta_{2j,\xi_i}$}
        (Yj);
  \path (xi) edge[dashed,bend right=25]
        node[pos=0.85, below=3pt] {$\delta_{1J,\xi_i},\,\delta_{2J,\xi_i}$}
        (YJ);
\end{tikzpicture}
\caption{Path diagram of the hybrid latent‐class IRT model. 
    Solid arrows represent the baseline proportional‐odds parameters 
    ($\tau_{jm},a_j$), and dashed arrows represent DIF effects 
    ($\delta_{1j,\xi_i},\delta_{2j,\xi_i}$).}
  \label{fig:hybrid‐latent‐class}
\end{figure}

Figure \ref{fig:hybrid‐latent‐class} illustrates the resulting hybrid latent-class IRT model. Solid arrows represent baseline proportional odds parameters ($\tau_{jm}$, $a_j$), while the dashed arrows indicate the class-specific DIF effects 
\((\delta_{1j,\xi_i}, \delta_{2j,\xi_i})\). 

\subsection{Marginal Likelihood Function}

The estimation for the proposed modeling framework is carried out using the marginal maximum likelihood function, where the latent variables are treated as random effects to be marginalised. The likelihood contribution for individual $i$ for the proposed model is:
\begin{equation}\label{eq:individual_likelihood}
L_i(\boldsymbol{\Theta}) = \sum_{k=0}^K \nu_k \int_{\mathbb{R}} \left(\prod_{j=1}^J P(Y_{ij} | \theta, k; \boldsymbol{\Theta})\right) \phi(\theta; \mu_k, \sigma_k^2) d\theta
\end{equation}
where $\phi(\cdot; \mu, \sigma^2)$ denotes the probability density function of the normal distribution with mean $\mu$ and variance $\sigma^2$. The parameter vector $\boldsymbol{\Theta}$ encompasses all unknown model parameters: threshold parameters $\{\tau_{jm}\}$, slope parameters $\{a_j\}$, DIF effects $\{\delta_{1jk}, \delta_{2jk}\}$, class probabilities $\{\nu_k\}$, and distributional parameters $\{\mu_k, \sigma_k^2\}$. The complete marginal log-likelihood function is given by:
\begin{equation}\label{eq:loglikelihood}
\ell(\boldsymbol{\Theta}) = \sum_{i=1}^N \log L_i(\boldsymbol{\Theta})
\end{equation}

\subsection{Regularised Estimation and Identifiability}
\label{sec:regularised-identifiability}

\subsubsection{Non-identifiability without anchor constraints}
\label{sec:nonidentifiability}
{
In the absence of known anchor items, the model is not identifiable without additional structural assumptions. 
The source of this non-identifiability can be seen by considering a non-reference class \(k \geq 1\). 
For individuals in class \(k\), the linear predictor in \eqref{eq:measurement_model} is
\[
\eta_{jkm}(\theta)
=
\tau_{jm} - (a_j+\delta_{2jk})\theta + \delta_{1jk}.
\]
Now consider a class-specific reparametrisation of the latent trait,
\[
\theta = b_k + c_k \theta^{*}, \qquad c_k>0.
\]
Under this transformation, the class-specific latent distribution becomes
\[
\theta^{*}\mid \xi_i=k
\sim
N\left(\frac{\mu_k-b_k}{c_k}, \frac{\sigma_k^2}{c_k^2}\right).
\]
At the same time, the linear predictor can be rewritten as
\[
\eta_{jkm}(\theta)
=
\tau_{jm}
-
c_k(a_j+\delta_{2jk})\theta^{*}
+
\{\delta_{1jk}-b_k(a_j+\delta_{2jk})\}.
\]
Thus, the same conditional response probabilities are obtained after defining
\[
\delta^{*}_{2jk}
=
c_k(a_j+\delta_{2jk})-a_j,
\qquad
\delta^{*}_{1jk}
=
\delta_{1jk}-b_k(a_j+\delta_{2jk}),
\]
and
\[
\mu^{*}_k=\frac{\mu_k-b_k}{c_k},
\qquad
\sigma^{*}_k=\frac{\sigma_k}{c_k}.
\]
The corresponding marginal likelihood is unchanged, because the transformation is simply a change of variable in the class-specific integral. 
Consequently, different combinations of 
\((\mu_k,\sigma_k,\delta_{1jk},\delta_{2jk})\) can yield equivalent likelihoods. 
A location shift \(b_k\) in the class-specific latent distribution can be absorbed by the uniform DIF parameters \(\delta_{1jk}\), while a scale transformation \(c_k\) can be absorbed by the non-uniform DIF parameters \(\delta_{2jk}\). 
Therefore, fixing \(\mu_0=0\), \(\sigma_0^2=1\), \(\delta_{1j0}=0\), and \(\delta_{2j0}=0\) identifies the reference-group scale, but does not by itself identify the class-specific latent distributions and DIF effects for the non-reference classes.
} {
Known anchor items would break this invariance by constraining selected DIF effects to zero. 
For example, if item \(j\) is known to be invariant in class \(k\), so that \(\delta_{1jk}=0\) and \(\delta_{2jk}=0\), then the transformed parameters must also satisfy \(\delta^{*}_{1jk}=0\) and \(\delta^{*}_{2jk}=0\). 
These restrictions constrain the admissible values of \(b_k\) and \(c_k\), thereby tying the class-specific latent distribution to the common measurement scale. 
However, in exploratory DIF analysis, such anchor items are typically unknown.}

\subsubsection{Regularisation}
\label{sec:sparse-regularisation}

To overcome this identifiability issue and simultaneously perform DIF detection, we adopt a sparsity assumption, namely that only a subset of items exhibit differential functioning across latent classes. 
This reflects many real applications, where DIF is expected to be rare or at least concentrated in a limited portion of the item set 
(see, e.g., \citealp{magis2015detection, tutz2015penalty, belzak2020improving, bauer2020simplifying, schauberger2020regularization}). 
The sparsity assumption is not only a way to perform variable selection; it also plays an identifying role. 
In the absence of known anchors, a regularised estimator favours, among the equivalent parametrisations described above, the solution in which most items are treated as invariant and only a limited number of item-by-class effects are attributed to DIF.

{The required degree of sparsity is problem dependent. 
In general, the approach is most appropriate when the number of DIF effects is small relative to the total number of possible class-specific DIF effects. 
Equivalently, for each non-reference class, there should be a sufficiently large set of approximately invariant items to define the common latent scale. 
If uniform or non-uniform DIF is pervasive across the item set, then the distinction between item-level DIF and genuine class differences in the latent trait distribution becomes weak. 
In such settings, the regularised estimator may still return a sparse solution, but the resulting decomposition of class differences into latent distributional differences and DIF effects should be interpreted with caution.}

We therefore impose an \(\ell_1\) penalty on the class-specific DIF parameters \(\{\delta_{1jk}, \delta_{2jk}\}\), which shrinks small effects toward zero and encourages a sparse DIF structure. 
Formally, the regularised estimator is defined as
\begin{equation}\label{eq:regularized_estimator}
\hat{\boldsymbol{\Theta}}^{(\lambda)} = \underset{\boldsymbol{\Theta} \in \Omega}{\arg\min} 
\left\{-\ell(\boldsymbol{\Theta}) + \lambda \mathcal{P}(\boldsymbol{\Theta})\right\},
\end{equation}
where \(\Omega\) denotes the constrained parameter space and 
\begin{equation}\label{eq:penalty}
\mathcal{P}(\boldsymbol{\Theta}) 
= \sum_{j=1}^J \sum_{k=1}^K \big(|\delta_{1jk}| + |\delta_{2jk}|\big)
\end{equation}
is the sparsity-inducing penalty. 
The tuning parameter \(\lambda \geq 0\) controls the degree of regularisation, with larger values enforcing stronger shrinkage and yielding fewer items flagged for DIF. {Equation \eqref{eq:regularized_estimator} uses a common tuning parameter for the uniform and non-uniform DIF effects. 
This, however, is not the only possible choice. A more general formulation would allow separate tuning parameters,
\[
\mathcal{P}_{\lambda_1,\lambda_2}(\boldsymbol{\Theta})
=
\lambda_1 \sum_{j=1}^J\sum_{k=1}^K |\delta_{1jk}|
+
\lambda_2 \sum_{j=1}^J\sum_{k=1}^K |\delta_{2jk}|,
\]
where \(\lambda_1\) controls sparsity in the uniform DIF effects and \(\lambda_2\) controls sparsity in the non-uniform DIF effects. 
This formulation may be useful when one type of DIF is expected to be more prevalent than the other, or when the two sets of parameters have different empirical scales. 
This distinction is relevant because \(\delta_{1jk}\) represents a location shift on the logit scale, whereas \(\delta_{2jk}\) modifies the item slope and affects response probabilities through its interaction with the latent trait. 
Consequently, a common penalty may induce different effective shrinkage on uniform and non-uniform DIF effects.}

{Another possible extension is an adaptive lasso penalty,
\[
\mathcal{P}_{\mathrm{ad}}(\boldsymbol{\Theta})
=
\sum_{j=1}^J\sum_{k=1}^K
\left(
w_{1jk}|\delta_{1jk}| + w_{2jk}|\delta_{2jk}|
\right),
\]
where \(w_{1jk}\) and \(w_{2jk}\) are data-dependent weights, for example obtained from an initial unpenalised or weakly penalised estimator. Such weights could reduce over-shrinkage of larger effects and partly account for differences in the empirical scales of uniform and non-uniform DIF parameters. In the present framework, we use a common \(\lambda\) to keep the tuning problem one-dimensional and computationally tractable, particularly because the model already involves latent class estimation, ordinal thresholds, and post-selection refitting. We nevertheless view separate tuning parameters and adaptive weights as useful extensions that fit within the proposed framework.} 

\subsection{Model Selection}

{Model selection involves both the choice of the regularisation parameter and the number of latent classes. 
Let \(C=K+1\) denote the total number of latent classes, and let \(\mathcal{C}\) denote the set of candidate class numbers considered. 
For each candidate value \(C \in \mathcal{C}\), we perform a grid search over a pre-specified set of tuning parameters \(\Lambda=\{\lambda^{(1)},\ldots,\lambda^{(M)}\}\}\).}

{For a fixed number of classes \(C\) and a candidate tuning parameter \(\lambda^{(m)}\), we first obtain the regularised estimate
\[
\hat{\boldsymbol{\Theta}}^{(C,\lambda^{(m)})}
=
\arg\min_{\boldsymbol{\Theta}\in\Omega_C}
\left\{
-\ell_C(\boldsymbol{\Theta})
+
\lambda^{(m)}\mathcal{P}(\boldsymbol{\Theta})
\right\},
\]
where \(\Omega_C\) denotes the parameter space for the model with \(C\) latent classes and \(\ell_C(\boldsymbol{\Theta})\) is the corresponding marginal log-likelihood. 
The regularised estimate determines the active sets of uniform and non-uniform DIF parameters,
\[
\mathcal{A}_1^{(C,\lambda^{(m)})}
=
\left\{(j,k):
|\hat{\delta}_{1jk}^{(C,\lambda^{(m)})}|>0
\right\},
\qquad
\mathcal{A}_2^{(C,\lambda^{(m)})}
=
\left\{(j,k):
|\hat{\delta}_{2jk}^{(C,\lambda^{(m)})}|>0
\right\}.
\]
We then refit the model without penalisation, holding the inactive DIF parameters fixed at zero.} This gives the confirmatory maximum likelihood estimate
\[
\tilde{\boldsymbol{\Theta}}^{(C,\lambda^{(m)})}
=
\arg\max_{\boldsymbol{\Theta}\in
\Omega_{\mathcal{A}_1^{(C,\lambda^{(m)})},\mathcal{A}_2^{(C,\lambda^{(m)})}}}
\ell_C(\boldsymbol{\Theta}).
\]
The BIC for the model defined by \(C\) and \(\lambda^{(m)}\) is computed from this confirmatory refit:
\begin{equation}\label{eq:BIC}
\mathrm{BIC}(C,\lambda^{(m)})
=
-2\ell_C\left(\tilde{\boldsymbol{\Theta}}^{(C,\lambda^{(m)})}\right)
+
\log(N)\,\mathrm{df}_{C,\lambda^{(m)}},
\end{equation}
where \(\mathrm{df}_{C,\lambda^{(m)}}\) is the number of freely estimated parameters in the refitted model, including the class proportions, class-specific distributional parameters, baseline item parameters, and the active uniform and non-uniform DIF parameters.

{For each fixed \(C\), the tuning parameter is selected as
\[
\hat{\lambda}_C
=
\arg\min_{\lambda^{(m)}\in\Lambda}
\mathrm{BIC}(C,\lambda^{(m)}).
\]
The resulting best post-selection model for each candidate number of classes is then compared using BIC, and the final number of classes is selected as
\[
\hat{C}
=
\arg\min_{C\in\mathcal{C}}
\mathrm{BIC}(C,\hat{\lambda}_C).
\]
Thus, the full model-selection procedure consists of: 
(i) fitting a regularised model over a grid of \(\lambda\) values for each candidate number of classes; 
(ii) selecting the active DIF structure induced by each \(\lambda\); 
(iii) refitting the corresponding confirmatory model without penalisation; 
(iv) selecting the best \(\lambda\) within each candidate class number using BIC; and 
(v) comparing the resulting confirmatory models across class numbers using BIC. 
Other model-selection criteria are available, but we use BIC because of its consistency properties \citep{shao1997asymptotic}}.

\subsection{Computational Algorithm}

The optimisation of the regularised objective function is accomplished through a modified EM algorithm that incorporates proximal gradient updates for handling the non-smooth penalty terms\footnote{The R code implementation is available here: \url{https://osf.io/xvu9y/overview?view_only=5d61ea6cce1b45e0b89415ec4888a49e}}. The algorithm alternates between an expectation step that computes posterior probabilities of latent class membership and latent trait values, and a maximisation step that employs proximal operators for the penalised parameter updates.

\textbf{Initialisation.} 
{Because the objective function is non-convex, the EM algorithm requires stable starting values. We use a data-driven initialisation strategy. A single-class model is first fitted without latent classes or DIF effects. The resulting estimates provide starting values for the common item parameters \(a_j\) and \(\tau_{jm}\). This initial fit places the item parameters on a common latent scale before class-specific distributional and DIF parameters are introduced.} {Class-specific starting values are then obtained by estimating individual latent trait scores under the preliminary model and applying a finite-mixture clustering procedure to these scores. The estimated mixing proportions initialise \(\nu_k\), while the cluster means and standard deviations initialise \(\mu_k\) and \(\sigma_k\) for the non-reference classes. The reference class is fixed at \(\mu_0=0\) and \(\sigma_0=1\) for identification. The non-reference class standard deviations are constrained to be positive and bounded away from zero.} {Lastly, the DIF parameters are initialised at or near zero. Small random perturbations around zero may be used for \(\delta_{1jk}\) and \(\delta_{2jk}\) to break symmetry across latent classes, while remaining consistent with the sparsity assumption that most items are initially treated as invariant. The DIF structure is then determined by the subsequent proximal-gradient updates and model-selection step.}

\textbf{E-step: Posterior computation.} Given current parameter estimates $\boldsymbol{\Theta}^{(t)}$, the expectation step computes the posterior distribution of the latent variables for each individual. For computational tractability, we approximate the integral in equation \eqref{eq:individual_likelihood} using Gauss-Hermite quadrature with $G$ nodes. Let $\{\theta_g, w_g\}_{g=1}^G$ denote the quadrature nodes and weights for the standard normal distribution. The posterior probability that individual $i$ belongs to latent class $k$ and has latent trait value $\theta_g$ is:
\begin{equation}\label{eq:posterior}
q_{ikg} = \frac{\nu_k w_g \prod_{j=1}^J P(Y_{ij} | \theta_g, k; \boldsymbol{\Theta}^{(t)}) \phi(\theta_g; \mu_k, \sigma_k^2)}{\sum_{k'=0}^K \sum_{g'=1}^G \nu_{k'} w_{g'} \prod_{j=1}^J P(Y_{ij} | \theta_{g'}, k'; \boldsymbol{\Theta}^{(t)}) \phi(\theta_{g'}; \mu_{k'}, \sigma_{k'}^2)}
\end{equation}
where $\phi(\theta_g; \mu_k, \sigma_k^2) = \sigma_k^{-1} \phi\left(\frac{\theta_g - \mu_k}{\sigma_k}\right)$ represents the density of the class-specific latent trait distribution evaluated at the transformed quadrature node. These posterior probabilities serve as weights in the subsequent maximisation step. {The posterior weights \(q_{ikg}^{(t)}\) define the expected complete-data log-likelihood used in the M-step:
\[
Q(\boldsymbol{\Theta}\mid\boldsymbol{\Theta}^{(t)})
=
\sum_{i=1}^{N}\sum_{k=0}^{K}\sum_{g=1}^{G}
q_{ikg}^{(t)}
\left[
\log \nu_k
+
\sum_{j=1}^{J}\log P(Y_{ij}\mid \theta_g,k;\boldsymbol{\Theta})
+
\log \phi(\theta_g;\mu_k,\sigma_k^2)
\right].
\]
This quantity is the Gauss--Hermite quadrature approximation to the expected complete-data log-likelihood under the current posterior distribution of the latent variables.}

\textbf{M-step: Parameter updates.} The maximisation step updates all model parameters by maximising \(Q(\boldsymbol{\Theta}\mid\boldsymbol{\Theta}^{(t)})\) subject to the \(\ell_1\) penalty. {The full marginal objective is not separable in all parameters, but the \(\ell_1\) penalty is separable across the DIF effects. We therefore use a block-coordinate optimisation strategy, updating each parameter block conditional on the current values of the others. This allows the penalised DIF parameters to be updated using proximal-gradient soft-thresholding, while the unpenalised parameters are updated by gradient descent steps. Because the M-step is not solved in closed form, but is instead approximated using block-coordinate gradient and proximal-gradient updates, the usual monotonicity property of an exact EM algorithm is not automatic. The backtracking line search is therefore used as a safeguard within each M-step to ensure that the proposed block updates produce descent in the full penalised objective,
\[
f(\boldsymbol{\Theta})
=
-\ell(\boldsymbol{\Theta})+\lambda\mathcal{P}(\boldsymbol{\Theta}).
\]
This issue is related to the block-coordinate implementation: each block update is computed conditional on the current values of the other parameter blocks, and a step size that is too large may decrease the local quadratic approximation but fail to improve the full penalised objective.}

{Given initial step sizes \(\{\eta_a^{(0)}, \eta_{\tau}^{(0)}, \eta_{\delta_1}^{(0)}, \eta_{\delta_2}^{(0)}, \eta_{\mu}^{(0)}, \eta_{\sigma}^{(0)}\}\) and a reduction factor \(\rho \in (0,1)\), we first compute tentative updates for all parameter blocks. 
If the resulting candidate \(\boldsymbol{\Theta}^{(t+1)}\) satisfies
\[
f(\boldsymbol{\Theta}^{(t+1)}) \leq f(\boldsymbol{\Theta}^{(t)}),
\]
the update is accepted. Otherwise, the step sizes are replaced by \(\rho\eta\), the tentative updates are recomputed, and the check is repeated until descent is achieved or a maximum number of backtracking steps is reached. Thus, backtracking is not introduced because the EM objective is intrinsically non-monotone, but because the M-step is implemented approximately rather than 
maximised exactly.}

{\textit{Note. Regularisation-based DIF detection has also been studied in multidimensional IRT models using lasso and adaptive lasso penalties. \cite{wang2023using} proposed an expectation--maximisation--maximisation (EMM) algorithm for identifying DIF in multidimensional 2PL models. Our procedure is related in that it uses penalisation within an EM-type algorithm for model selection, but differs in the role assigned to the penalised estimator. In the proposed approach, the penalised estimator is used only to identify the active set of DIF parameters. Once this active set has been selected, the corresponding parameters are re-estimated without penalisation under the constrained model, with the aim of reducing the shrinkage bias induced by the \(\ell_1\) penalty.}}

The resulting block updates are as follows. For the class probability parameter, the M-step yields a closed-form update:
\begin{equation}\label{eq:pi_update}
\nu_k^{(t+1)} = \frac{1}{N} \sum_{i=1}^N \sum_{g=1}^G q_{ikg}
\end{equation}

For the distributional parameters of the non-reference classes ($k \geq 1$), we compute the gradient of the expected complete-data log-likelihood and apply gradient descent updates:
\begin{equation}\label{eq:mu_sigma_update}
\mu_k^{(t+1)} = \mu_k^{(t)} - \eta_{\mu} \nabla_{\mu_k} Q(\boldsymbol{\Theta} | \boldsymbol{\Theta}^{(t)}), \quad
\sigma_k^{(t+1)} = \sigma_k^{(t)} - \eta_{\sigma} \nabla_{\sigma_k} Q(\boldsymbol{\Theta} | \boldsymbol{\Theta}^{(t)})
\end{equation}
where $\eta_{\mu}$ and $\eta_{\sigma}$ denote step sizes, chosen by a line search procedure.

The discrimination parameters and thresholds, which are not subject to penalisation, are also updated via standard gradient descent:
\begin{equation}\label{eq:a_d_update}
a_j^{(t+1)} = a_j^{(t)} - \eta_a \nabla_{a_j} Q(\boldsymbol{\Theta} | \boldsymbol{\Theta}^{(t)}), \quad
\tau_{jm}^{(t+1)} = \tau_{jm}^{(t)} - \eta_{\tau} \nabla_{\tau_{jm}} Q(\boldsymbol{\Theta} | \boldsymbol{\Theta}^{(t)})
\end{equation}

The DIF effect parameters, which are subject to the $\ell_1$ penalty, are updated using proximal gradient descent. For the uniform DIF parameters:
\begin{equation*}\label{eq:delta1_update}
\delta_{1jk}^{(t+1)} = 
\text{prox}_{\lambda \eta_{\delta_1} |\cdot|}
\left(
\delta_{1jk}^{(t)} - \eta_{\delta_1}
\nabla_{\delta_{1jk}} Q(\boldsymbol{\Theta} \mid \boldsymbol{\Theta}^{(t)})
\right).
\end{equation*}
and similarly for the non-uniform DIF parameters $\delta_{2jk}$: 
$$
\delta_{2jk}^{(t+1)} = 
\text{prox}_{\lambda \eta_{\delta_2} |\cdot|}
\left(
\delta_{2jk}^{(t)} - \eta_{\delta_2}
\nabla_{\delta_{2jk}} Q(\boldsymbol{\Theta} \mid \boldsymbol{\Theta}^{(t)})
\right).    
$$

The proximal operator for the $\ell_1$ penalty admits a closed-form solution via soft-thresholding:
\begin{equation}\label{eq:proximal_operator}
\text{prox}_{\lambda|\cdot|}(x) = \text{sign}(x) \max(|x| - \lambda, 0)
\end{equation}
ensuring computational efficiency while maintaining the exact sparsity-inducing properties of the penalty function.

The EM algorithm iterates between the E-step and M-step until convergence, which is declared when the absolute change in the penalised objective function falls below a pre-specified tolerance $|f^{(t+1)} - f^{(t)}| < \epsilon_{EM}$.

\subsubsection{Confirmatory maximum likelihood estimation.} Following the selection of the optimal tuning parameter \(\hat{\lambda}\) via BIC, we identify the active sets of uniform and non-uniform DIF parameters as
\[
\mathcal{A}_1 =
\left\{(j,k):
|\hat{\delta}_{1jk}^{(\hat{\lambda})}| > 0
\right\},
\qquad
\mathcal{A}_2 =
\left\{(j,k):
|\hat{\delta}_{2jk}^{(\hat{\lambda})}| > 0
\right\}.
\] We then compute the constrained maximum likelihood estimator by solving:
\begin{equation}\label{eq:confirmatory_mle}
\tilde{\boldsymbol{\Theta}}^{(\hat{\lambda})} =
\arg\max_{\boldsymbol{\Theta} \in \Omega_{\mathcal{A}_1,\mathcal{A}_2}}
\ell(\boldsymbol{\Theta}).
\end{equation}
where
\[
\Omega_{\mathcal{A}_1,\mathcal{A}_2}
=
\left\{
\boldsymbol{\Theta} \in \Omega:
\delta_{1jk}=0 \text{ for all } (j,k)\notin\mathcal{A}_1,
\quad
\delta_{2jk}=0 \text{ for all } (j,k)\notin\mathcal{A}_2
\right\}
\]
represents the parameter space with zero constraints imposed on the inactive uniform and non-uniform DIF parameters. This constrained optimisation is executed using the same EM algorithm as described above, but with $\lambda = 0$ and the inactive parameters held fixed at zero throughout all iterations. The resulting estimator $\tilde{\boldsymbol{\Theta}}^{(\hat{\lambda})}$ retains the sparse structure identified during model selection whilst eliminating the shrinkage bias induced by the penalty.

{This post-selection refit is used to improve interpretability of the estimated effect sizes, but we do not report standard errors or conduct formal hypothesis tests for the selected DIF parameters. Valid inference in this setting is non-trivial because the active DIF structure is selected from the data, the likelihood contains latent classes, and the regularised estimation step introduces non-smooth model-selection uncertainty. Consequently, conventional standard errors from the refitted model would not account for uncertainty in the selection of the sparse DIF structure or the number of latent classes.}


\section{Empirical Study}\label{sec:Empirical}

\subsection{Data}

We illustrate the proposed methodology using data from the Fisher Temperament Inventory (FTI) \citep{brown2013neural}, a theory-driven personality measure grounded in neurobiological models of temperament. The FTI was developed to measure four suites of behavioral traits, which is hypothesized to be associated with four broad neural systems: dopamine/norepinephrine (Curious/Energetic), serotonin (Cautious/Social Norm Compliant), testosterone (Analytical/Tough-minded), and estrogen/oxytocin (Prosocial/Empathetic). {The following analysis is intended to illustrate the proposed workflow and the kind of measurement heterogeneity the method can surface on real data, rather than to establish that a particular substantively meaningful subgroup exists in this instrument. Labelling the recovered latent class is not the objective: the method's recovery properties are established in the simulation study of Section 4, where the latent classes are unlabelled by construction, while the present example shows what the procedure yields in practice.}

The test was administered through openpsychometrics.org (\url{https://openpsychometrics.org/tests/FTI/}), with responses from $N = 4{,}967$ individuals. For illustration purposes, we focus on the Analytical/Tough-minded dimension. This subscale comprises 14 items, each measured on a four-point Likert scale (1=Strongly Disagree, 2=Disagree, 3=Agree, 4=Strongly Agree). Respondents indicate their agreement with statements such as:

\begin{itemize}
    \item I am able to solve problems without letting emotion get in the way.
    \item I am tough-minded.
    \item I think it is important to be direct.
    \item When making a decision, I like to stick to the facts rather than letting emotions influence the decisions.
\end{itemize}

The wordings of all items are provided in the supplementary material.

\subsection{Model fit and class structure}

We fitted the proposed model with one, two, and three latent classes. Model selection was based on the BIC, with results summarised in Table~\ref{tab:model-fit}.

\begin{table}[htbp]
\centering
\caption{Model comparison for the analysed FTI subscale using BIC.}
\label{tab:model-fit}
\begin{tabular}{lccc}
\hline
Fit index & 1-class & 2-class & 3-class \\ \hline
BIC               & 158{,}737.4 & 155{,}673.7 & 155{,}724.3 \\ \hline
\end{tabular}
\end{table}

The two-class model yields the lowest BIC ($\mathrm{BIC}=155{,}673.7$) and is therefore selected for the subsequent analyses. In this specification, the first class serves as the reference class, with the latent trait standardised to have mean $0$ and variance $1$. The second class has estimated latent trait mean $\mu_2 = -0.442$ and variance $\sigma_2 = 0.769$, indicating lower average levels and reduced variability on the latent dimension relative to the reference class. The estimated class proportion for the second class is
$
\pi_2 = 0.109,
$
so that approximately $10.9\%$ of respondents belong to this class and the remaining $89.1\%$ belong to the reference class.

\subsection{Item parameter estimates and DIF}

Table~\ref{tab:item-params} displays the estimated item discrimination parameters together with the estimated class-specific intercept and slope DIF effects. Several items exhibit clear DIF between the two latent classes. Notably, item 1 (\emph{``I understand complex machines easily''}), item 2 (\emph{``I enjoy competitive conversations''}), item 8 (\emph{``I am tough-minded''}), item 9 (\emph{``Debating is a good way to match my wits with others''}), and item 12 (\emph{``I like to avoid the nuances and say exactly what I mean''}) show the largest differences. We note that items exhibiting uniform DIF tend to be self-descriptive statements, whereas items with non-uniform DIF involve socially situated behaviours, suggesting distinct response mechanisms underlying the two forms of non-invariance.

Item 3 (\emph{``I am intrigued by rules and patterns that govern systems''}), item 5 (\emph{``I pursue intellectual topics thoroughly and regularly''}), item 13 (\emph{``I think it is important to be direct''}), and item 14 (\emph{``When making a decision, I like to stick to the facts rather than be swayed by people''}) demonstrate no DIF.

\begin{table}[htbp]
\centering
\caption{Estimated item discrimination and DIF parameters for the two-class model on the Analytical/Tough-minded subscale. Intercept and slope DIF values reflect post-threshold estimates.}
\label{tab:item-params}
\begin{tabular}{lccc}
\hline
Item & Discrimination $\hat{a}_j$ & Intercept DIF $\hat{\delta}_{1j}$ & Slope DIF $\hat{\delta}_{2j}$ \\ \hline
1  & 1.003 & -0.094 & 0.000 \\
2  & 1.760 &  0.000 & -0.259 \\
3  & 0.889 &  0.000 &  0.000 \\
4  & 1.366 & -0.086 & -0.116 \\
5  & 0.883 &  0.000 &  0.000 \\
6  & 1.554 &  0.000 & -0.089 \\
7  & 1.341 &  0.000 & -0.170 \\
8  & 1.014 & -0.161 &  0.000 \\
9  & 0.877 & -0.096 & -0.171 \\
10 & 0.642 &  0.000 & -0.129 \\
11 & 0.889 &  0.000 & -0.115 \\
12 & 0.790 &  0.000 & -0.235 \\
13 & 0.988 &  0.000 &  0.000 \\
14 & 1.597 &  0.000 &  0.000 \\ \hline
\end{tabular}
\end{table}


Table~\ref{tab:thresholds} reports the estimated threshold parameters $\tau_{jm}$. Each row corresponds to a single item, and each column represents the estimated latent trait location required to endorse response category $m$ or higher (for $m=2,3,4$). More negative values indicate a lower required level on the latent dimension to choose higher response options, while larger thresholds correspond to a higher required endorsement level. Across items, the first threshold is generally large and negative (typically between $-2$ and $-5$), indicating that respondents require only a modest latent level to move above the lowest response category. The second threshold is closer to zero for most items, suggesting noticeably greater latent trait level is needed to move from a low to a moderate endorsement level. The third threshold is consistently positive (often between $1$ and $2.5$), reflecting that endorsing the highest agreement category requires a substantially higher position on the latent trait.

\begin{table}[htbp]
\centering
\caption{Estimated threshold parameters $\tau_{jm}$ for the analysed FTI items. Thresholds represent latent trait values required to endorse category $m$ or higher.}
\label{tab:thresholds}
\begin{tabular}{lccc}
\hline
Item & $\tau_{j1}$ & $\tau_{j2}$ & $\tau_{j3}$ \\ \hline
1 & -2.366 &  0.005 & 2.456 \\
2 & -2.254 & -0.470 & 1.540 \\
3 & -3.061 & -0.958 & 1.395 \\
4 & -4.393 & -1.635 & 1.356 \\
5 & -4.823 & -2.074 & 0.428 \\
6 & -3.577 & -0.787 & 2.031 \\
7 & -4.786 & -2.334 & 0.529 \\
8 & -3.426 & -0.973 & 1.535 \\
9 & -2.631 & -0.617 & 1.490 \\
10 & -1.375 &  0.608 & 2.376 \\
11 & -2.330 & -0.286 & 1.531 \\
12 & -3.095 & -0.492 & 1.573 \\
13 & -4.933 & -2.158 & 1.107 \\
14 & -4.042 & -0.758 & 1.835 \\ \hline
\end{tabular}
\end{table}

\subsection{Post-hoc class characterisation}
{Because the latent classes are inferred from the response data, they should not be assumed to correspond directly to observed demographic groups. We therefore use available background variables only as an external characterisation of the estimated classes, not as variables that define the comparison groups. The purpose of this analysis is to assess whether class membership is associated with observed respondent characteristics and to aid interpretation of the latent structure. These associations should be viewed as descriptive and exploratory.} Rather than assigning each respondent to a class using a hard cut-off, we worked with the individual posterior probabilities
\[
p_i = P(\text{Class 2} \mid \text{responses}_i),
\]
which provide a continuous measure of each respondent's propensity to belong to Class~2. This ``soft'' representation is particularly appropriate here, as the estimated size of Class~2 is relatively small ($\hat{\pi}_2 \approx 0.11$), and hard classification would assign many respondents to the reference class with substantial uncertainty.

We considered age, family size, education, urban/rural residence, gender, native English status, handedness, religion, sexual orientation, race/ethnicity, voting status, and marital status. For the two continuous variables, age and family size, we compared the distributions across the two latent classes using two-sample \emph{t}-tests. The mean age was $31.65$ years in Class~1 and $32.40$ years in Class~2, with a borderline non-significant difference ($t=-1.92$, $p=0.054$). Similarly, mean family size was $2.52$ in Class~1 and $2.60$ in Class~2, again with a $p$-value close to the significance level ($t=-1.93$, $p=0.054$).

For categorical covariates, we fitted contingency tables of latent class by covariate level and used chi-square tests to assess association. Education and urban status showed no evidence of class-related differences (education: $\chi^2(3)=0.87$, $p=0.83$; urban/rural residence: $\chi^2(2)=0.68$, $p=0.71$). Likewise, there was no strong evidence of association for religion, sexual orientation, race/ethnicity, voting behaviour, or marital status (all $p>0.09$). In contrast, some background variables displayed statistically detectable differences. Native English speakers were more likely to belong to Class~2 than non-native speakers ($\chi^2(1)=9.78$, $p=0.0018$), and handedness also showed a significant association with class membership ($\chi^2(2)=17.54$, $p<0.001$).

Gender emerged as the most interpretable and strong correlate of latent class membership. A chi-square test of the class-by-gender table (Female/Male/Other) indicated a clear association ($\chi^2(2)=150.28$, $p<0.001$). To make use of the soft class information, we also compared the distributions of the posterior probabilities $p_i$ across gender groups. A Kruskal--Wallis test confirmed differences in $P(\text{Class 2} \mid \text{responses})$ by gender ($\chi^2(2)=200.2$, $p<0.001$), and a Wilcoxon rank-sum test restricted to male versus female respondents indicated a highly significant shift in location ($p<0.001$). {Note that the detected DIF should not be interpreted as DIF with respect to gender, native language, or handedness. These variables were not used to define the latent classes, and their associations with class membership were examined only after model estimation. Thus, the flagged items indicate class-specific DIF under the fitted latent mixture model, while the background-variable analysis provides only partial descriptive context for the latent classes.}

We also compared the baseline one-class model to evaluate how gender-related differences manifest when mixture structure is omitted. In the one-class IRT model, females showed clearly higher latent trait estimates than males (t=22.2, p<.001; difference $\approx$ 0.53 SD). This large mean shift is consistent with the idea that a single-population model compresses gender differences into the latent scale, whereas the two-class model instead captures them as differences in class membership probabilities.

These results suggest that the latent classes are not strongly structured by standard socio-demographic variables: age, education, and most background characteristics are very similar across classes. Gender, and to a lesser extent native language and handedness, appear to be associated with the probability of belonging to the smaller class, but the effect sizes are rather small. Thus, the classes identified by the model should not be interpreted as demographic groups; rather, the covariate analysis illustrates how auxiliary variables can provide clues about the behavioural or temperament-related differences captured by the latent structure. Ultimately, the goal is to obtain higher-quality measurements of the latent trait, adjusting for potential group-related differences. Taken together, these findings are consistent with the mixture model as a flexible model-based representation of heterogeneity in the data.

In this context, it is useful to distinguish between the two types of DIF identified by the model. Items exhibiting primarily uniform DIF tend to be broad self-descriptive statements, whereas items displaying non-uniform DIF more often involve socially situated behaviours. While these patterns are exploratory, they suggest that the latent classes may capture subtle differences in how respondents translate underlying dispositions into expressed behaviour. The associations with gender, native language, and handedness therefore provide contextual cues for interpretation, but the latent structure remains defined by the response model rather than by these background variables.

\section{Simulation Study}\label{sec:Simulation}

To evaluate the finite-sample performance of the proposed methodology, we conducted a simulation study examining parameter recovery, classification accuracy, and DIF detection performance under varying conditions. The study considers both a two-class and a three-class implementation, systematically varying sample size, number of items, and class proportion to assess the method's behaviour across several scenarios. For each condition, we use 100 replications.

\subsection{Simulation Design}

\textbf{Data generating mechanism.} We simulate ordinal response data from the proportional odds model specified in equation \eqref{eq:measurement_model} with $K = \{1, 2\}$ latent classes (note that $K=1$ means two latent classes, and $K=2$ means three latent classes). Each simulated data set comprises $N=\{1,000, 5,000\}$ respondents answering $J = \{15, 25, 50\}$ items, each with $M_j = 4$ ordered response categories. For the two-class case, the latent class membership follows a Bernoulli distribution with $\xi_i \sim \text{Bernoulli}(\pi)$, where $\pi=\{0.1, 0.3, 0.5\}$ represents the proportion of respondents in the focal group. Conditional on class membership, the latent traits are drawn from $\theta_i | \xi_i = 0 \sim \mathcal{N}(0, 1)$ for the reference group and $\theta_i | \xi_i = 1 \sim \mathcal{N}(\mu_1, \sigma_1^2)$ for the focal group, with $\mu_1 = 1$ and $\sigma_1 = 0.8$ in the two-group case, and additionally $\mu_2 = 1.5$ and $\sigma_2 = 0.75$ in the three-group case. For the three-group case, the mixing proportions are fixed to $\pi_0=0.5$, $\pi_1 = 0.3$, and $\pi_2 = 0.2$.

The item parameters are generated as follows. Baseline discrimination parameters are drawn independently from a uniform distribution, $a_j \sim \text{Uniform}(0.5, 1.5)$ for $j = 1, \ldots, J$. For each item, three threshold parameters are generated by first sampling from $\text{Uniform}(-2, 2)$ and then imposing the ordering constraint $\tau_{j1} < \tau_{j2} < \tau_{j3}$. We introduce DIF by specifying 5, 10, and 20 DIF items for $J=15$, $J=25$, and $J=50$, respectively; each DIF item is modeled to exhibit both uniform and non-uniform DIF. The uniform DIF effects are drawn from $\delta_{1j} \sim \text{Uniform}(1, 1.5)$ for DIF items and set to zero otherwise. Similarly, non-uniform DIF effects are generated as $\delta_{2j} \sim \text{Uniform}(0.5, 1)$ for the same $p$ items, with $\delta_{2j} = 0$ for the remaining $J - p$ anchor items. In the three-group case, the uniform and non-uniform DIF parameters for the third class were generated from a uniform distribution on the interval (0.5, 1).

\textbf{Estimation procedure.} For each simulated dataset, we apply the penalised EM algorithm described in Section \ref{sec:statistical-framework}. The tuning parameter $\lambda$ is selected from a grid of 10 values, $\lambda \in \{10^{-6}, 10^{-5.78}, \ldots, 10^{-2}\}$, via BIC as specified in equation \eqref{eq:BIC}. Following the BIC-based selection of $\lambda$, parameters flagged as non-zero by the penalised estimator are retained whilst those shrunken to 0 are set to zero. The constrained maximum likelihood estimator is then computed on this selected model to obtain final parameter estimates and calculate the performance metrics.

To provide an upper bound on classification performance, we also fit an oracle model for each simulated dataset. The oracle model uses the true item parameters $\{a_j, \tau_{jm}, \delta_{1j}, \delta_{2j}\}$ and true distributional parameters $\{\pi, \mu, \sigma\}$ to compute posterior class membership probabilities via equation \eqref{eq:individual_likelihood}. Respondents are assigned to the class with the highest posterior probability, and classification accuracy is computed by comparing these assignments to the true class labels. This oracle performance establishes the theoretical limit of classification accuracy given the data-generating mechanism and provides a benchmark for evaluating the proposed method's performance.

\textbf{Performance metrics.} We evaluate the method by the parameter estimation accuracy, respondent classification performance, and DIF detection accuracy. For parameter estimation, we compute the bias and root mean squared error (RMSE) for each parameter across the 100 replications within each condition. Item-level parameters ($a_j$, $\tau_{jm}$, $\delta_{1j}$, $\delta_{2j}$) are summarised by their RMSE to assess parameter recovery, whilst structural parameters ($\pi$, $\mu$, $\sigma$) are summarised by both bias and RMSE.

Respondent classification is assessed using two metrics. The classification error rate is computed as the proportion of respondents assigned to the incorrect latent class based on maximum a posteriori (MAP) assignment. The area under the receiver operating characteristic (ROC) curve (AUC) provides a threshold-invariant measure, with values approaching 1 indicating near-perfect separation between classes.

DIF detection performance is evaluated by the sensitivity and specificity at the item level. The true positive rate (TPR) measures the proportion of truly DIF-exhibiting items that are correctly identified by the method, whilst the false positive rate (FPR) quantifies the proportion of truly invariant anchor items incorrectly flagged as exhibiting DIF. These metrics are computed separately for uniform DIF ($\delta$) and non-uniform DIF ($\delta_2$) effects.

\subsection{Results}

\begin{figure}
    \centering
    \includegraphics[width=1\linewidth]{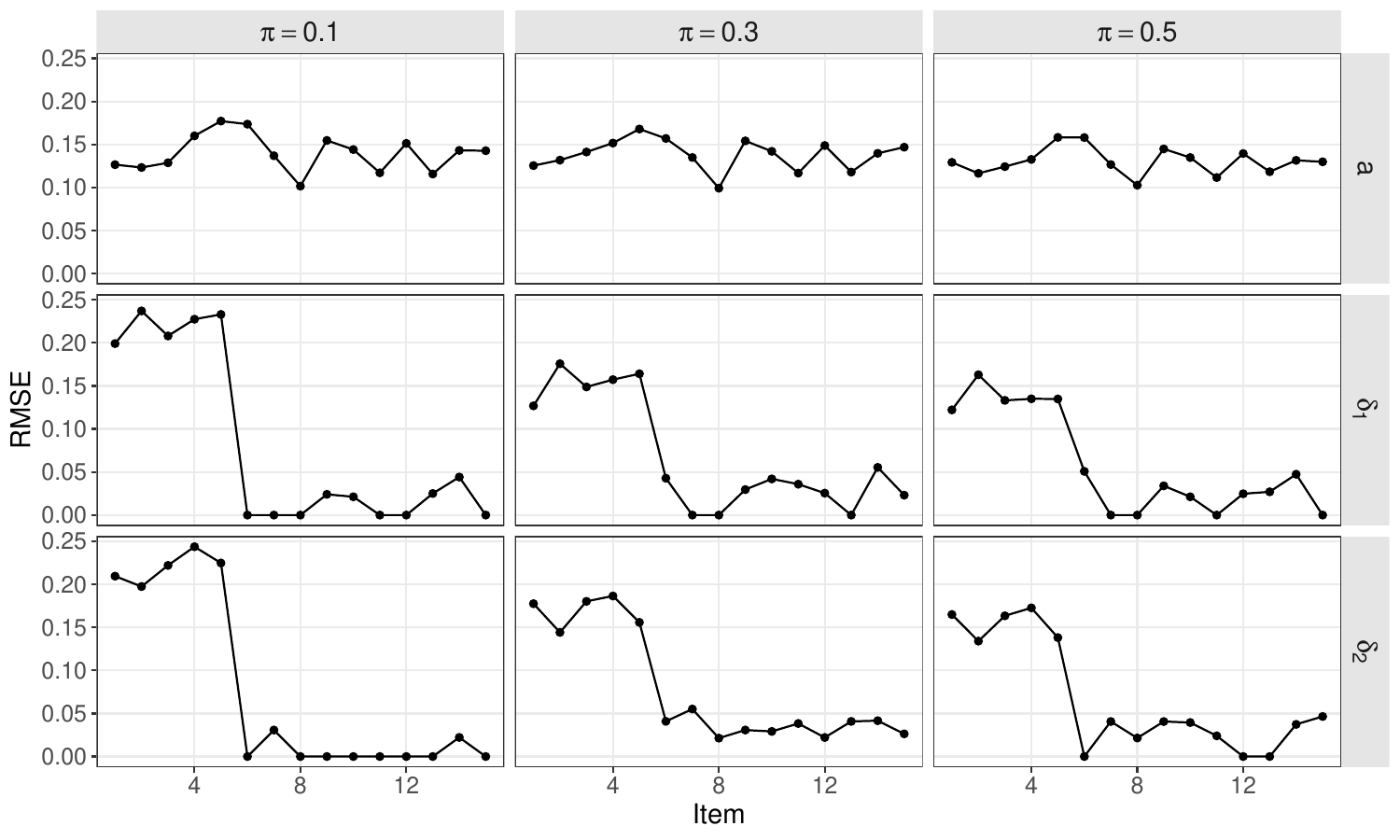}
    \caption{RMSE of item parameter estimates for $N=1,000$ and $J=15$ under the two-class case.}
    \label{fig:rmse_slope_dif_J15_N1000}
\end{figure}

\begin{figure}
    \centering
    \includegraphics[width=1\linewidth]{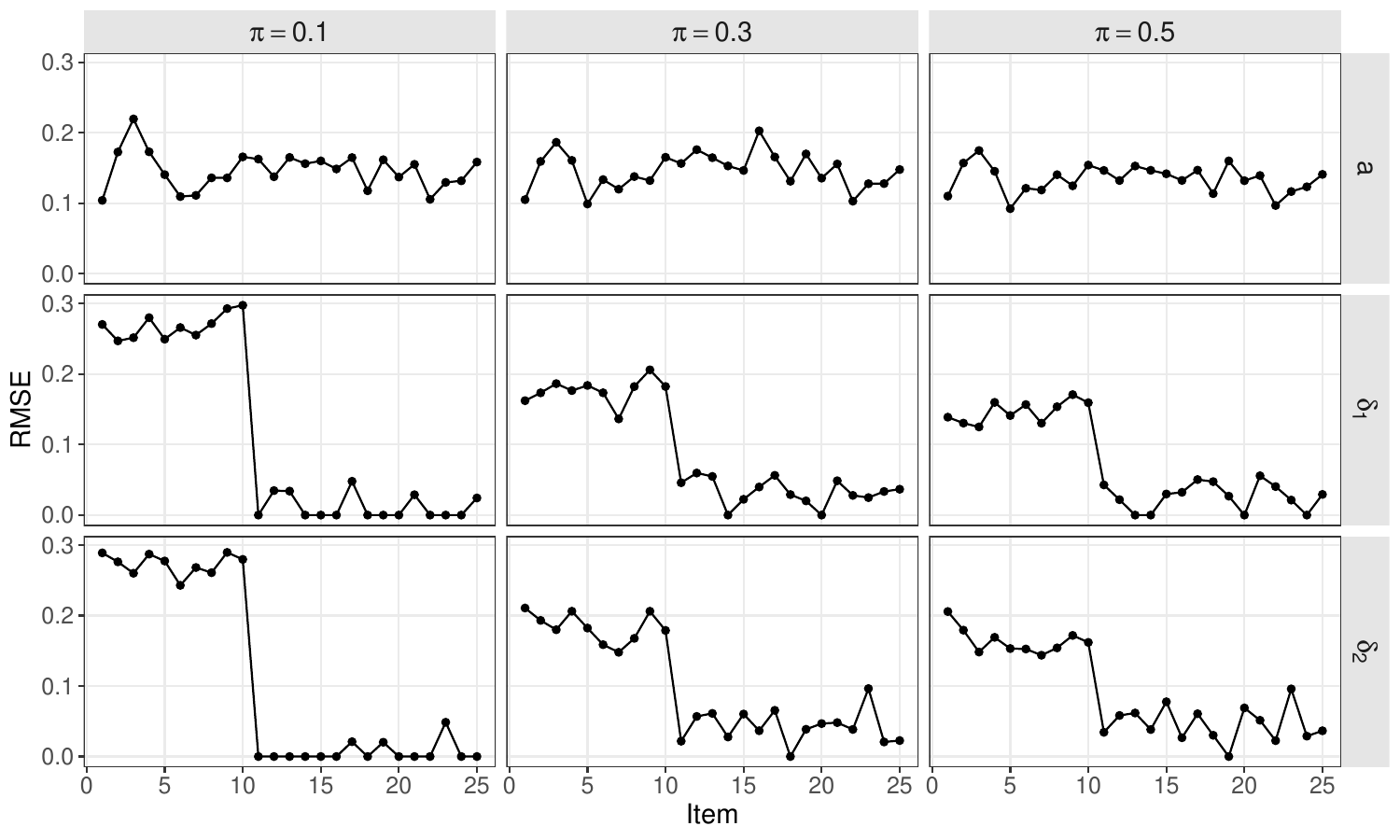}
    \caption{RMSE of item parameter estimates for $N=1,000$ and $J=25$ under the two-class case.}
    \label{fig:rmse_slope_dif_J25_N1000}
\end{figure}

\begin{figure}
    \centering
    \includegraphics[width=1\linewidth]{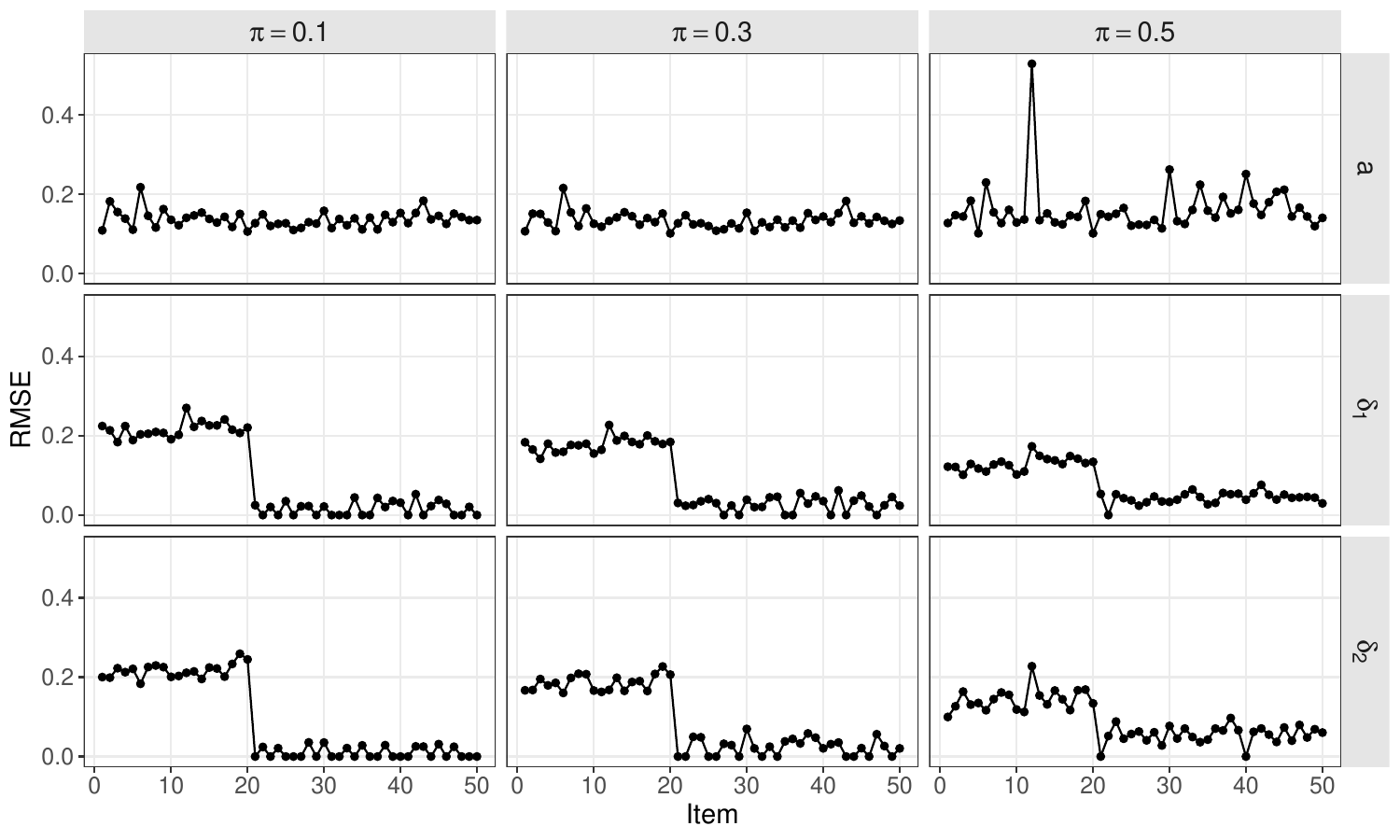}
    \caption{RMSE of item parameter estimates for $N=1,000$ and $J=50$ under the two-class case.}
    \label{fig:rmse_slope_dif_J50_N1000}
\end{figure}

\begin{figure}
    \centering
\includegraphics[width=1\linewidth]{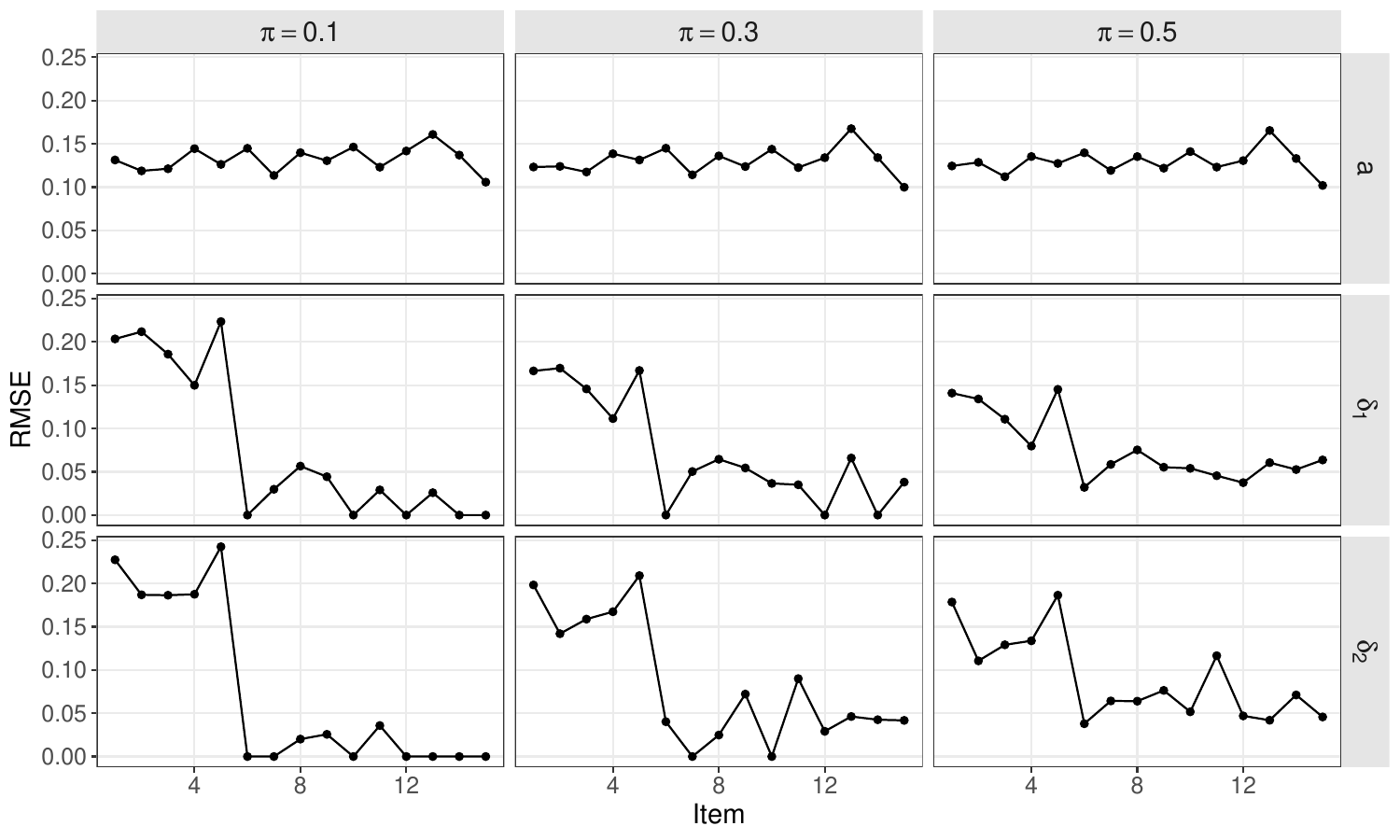}
    \caption{RMSE of item parameter estimates for $N=5,000$ and $J=15$ under the two-class case.}
\label{fig:rmse_slope_dif_J15_N5000}
\end{figure}

\begin{figure}
    \centering
\includegraphics[width=1\linewidth]{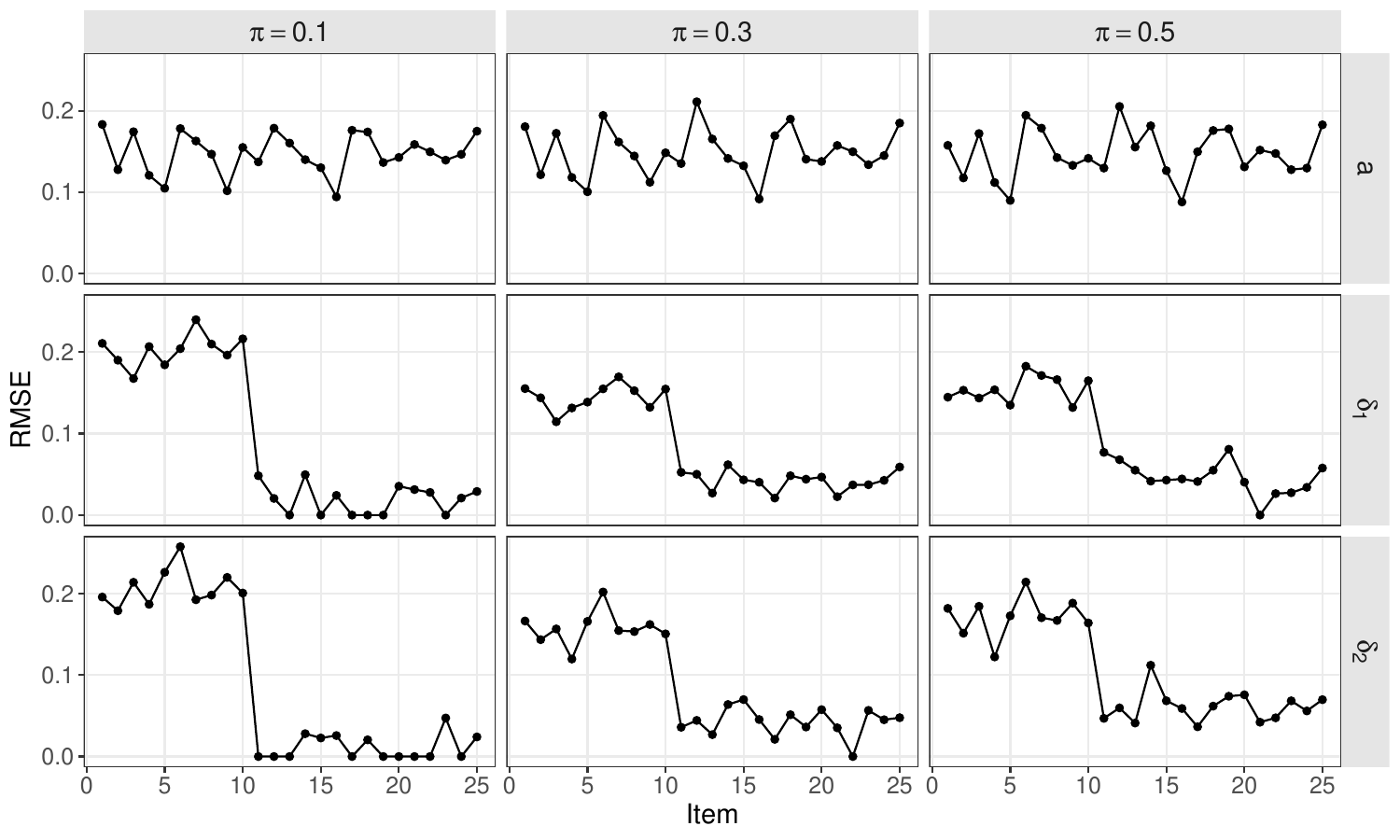}
    \caption{RMSE of item parameter estimates for $N=5,000$ and $J=25$ under the two-class case.}
\label{fig:rmse_slope_dif_J25_N5000}
\end{figure}

\begin{figure}
    \centering
\includegraphics[width=1\linewidth]{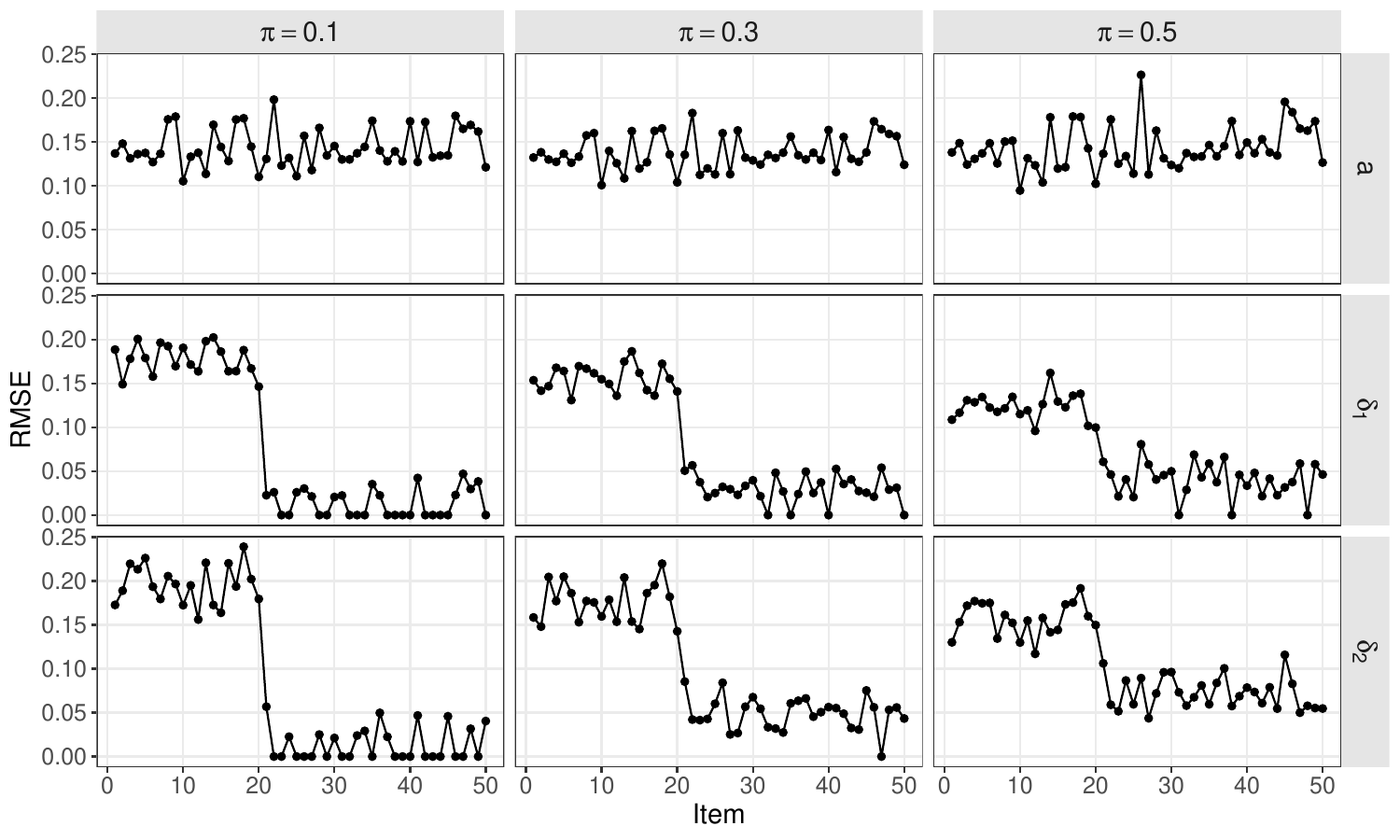}
    \caption{RMSE of item parameter estimates for $N=5,000$ and $J=50$ under the two-class case.}
\label{fig:rmse_slope_dif_J50_N5000}
\end{figure}

Figures~\ref{fig:rmse_slope_dif_J15_N1000}–\ref{fig:rmse_slope_dif_J50_N1000} display the RMSE for the item slope parameters $a_j$ and the DIF effects \(\delta_{1j}\) and \(\delta_{2j}\) for \(N = 1{,}000\) under the two-class model. Across all values of \(\pi\), RMSEs for the slope parameters are generally moderate and fairly homogeneous across items, with only a slight increase when the number of items grows. RMSEs for \(\delta_1\) and \(\delta_2\) are larger for items with nonzero DIF, while they are close to zero for items without DIF. As the two classes get closer in size, the RMSE curves generally reduce in magnitude, leading to more stable recovery of both slopes and DIF effects. The corresponding results for \(N = 5{,}000\) in Figures~\ref{fig:rmse_slope_dif_J15_N5000}–\ref{fig:rmse_slope_dif_J50_N5000} show a very similar pattern. Larger sample size leads to slightly smaller RMSEs and as the proportion of respondents in the latent classes even out, the RMSE curves somewhat decrease. 

\begin{figure}
    \centering
\includegraphics[width=1\linewidth]{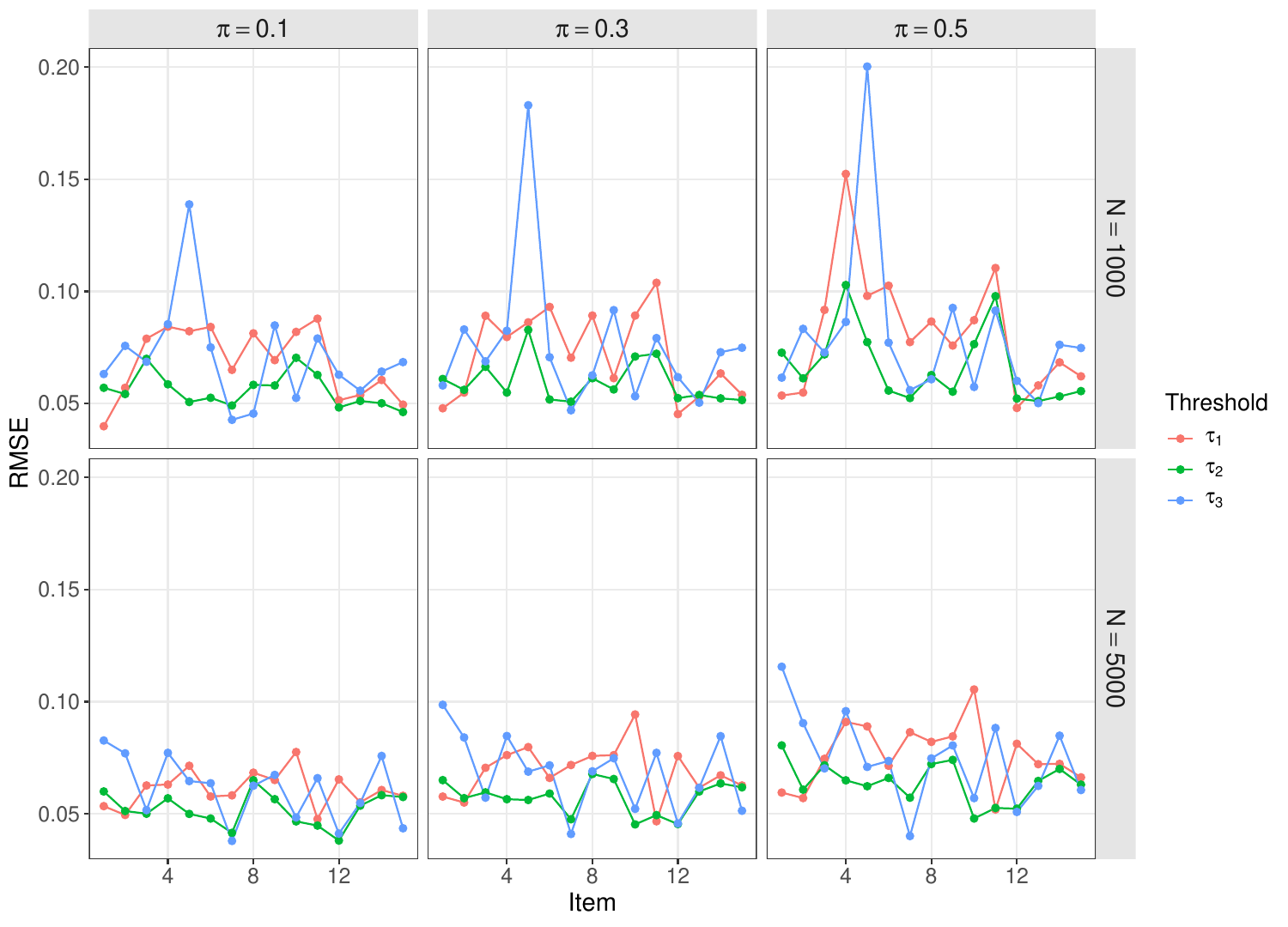}
    \caption{RMSE of threshold parameter estimates for $J=15$, $N=1000$ and $N=5000$ under the two-class case.}
\label{fig:rmse_thresholds_J15_N100050000}
\end{figure}

\begin{figure}
    \centering
\includegraphics[width=1\linewidth]{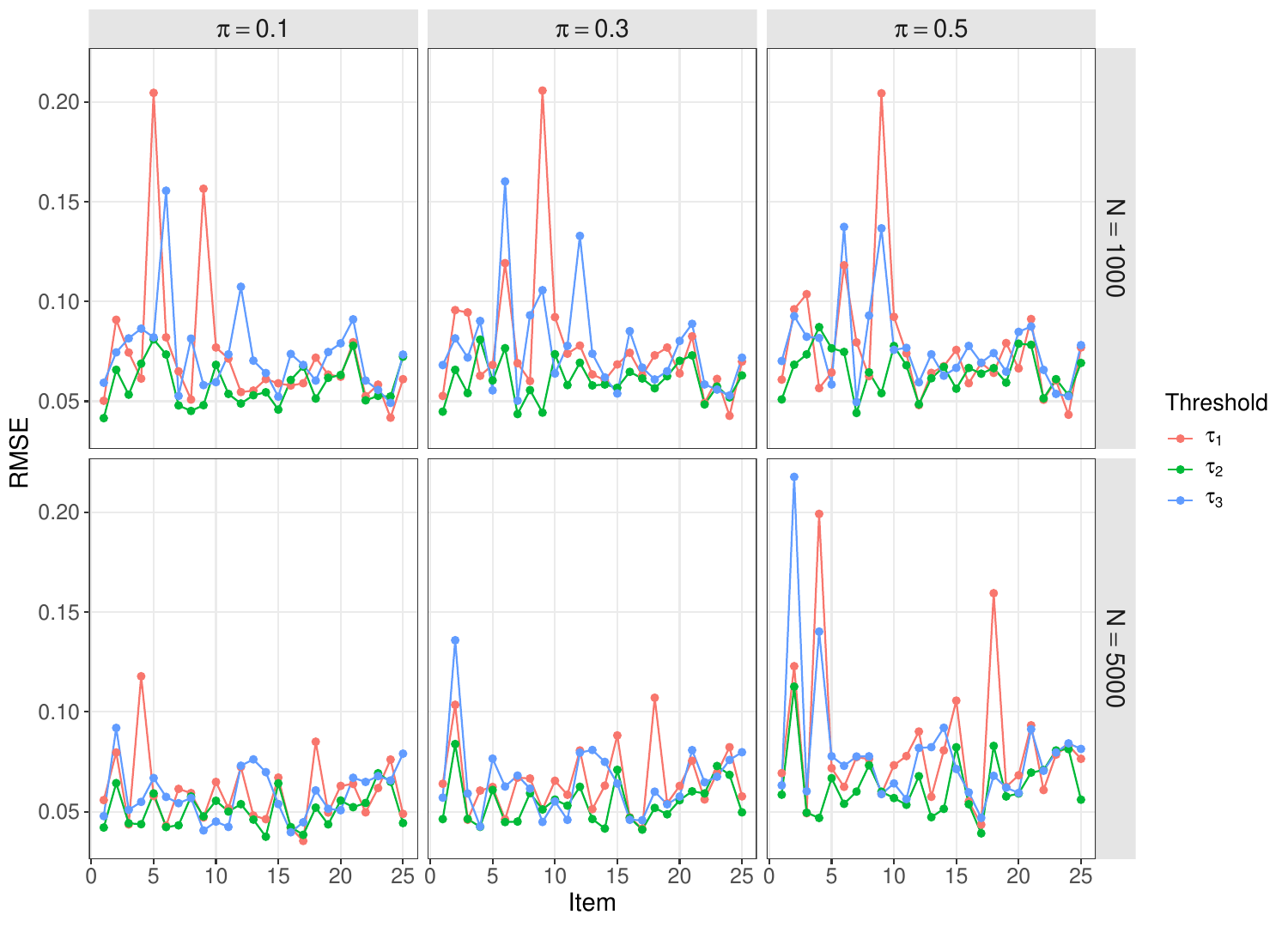}
    \caption{RMSE of threshold parameter estimates for $J=25$, $N=1000$ and $N=5000$ under the two-class case.}
\label{fig:rmse_thresholds_J25_N100050000}
\end{figure}

\begin{figure}
    \centering
\includegraphics[width=1\linewidth]{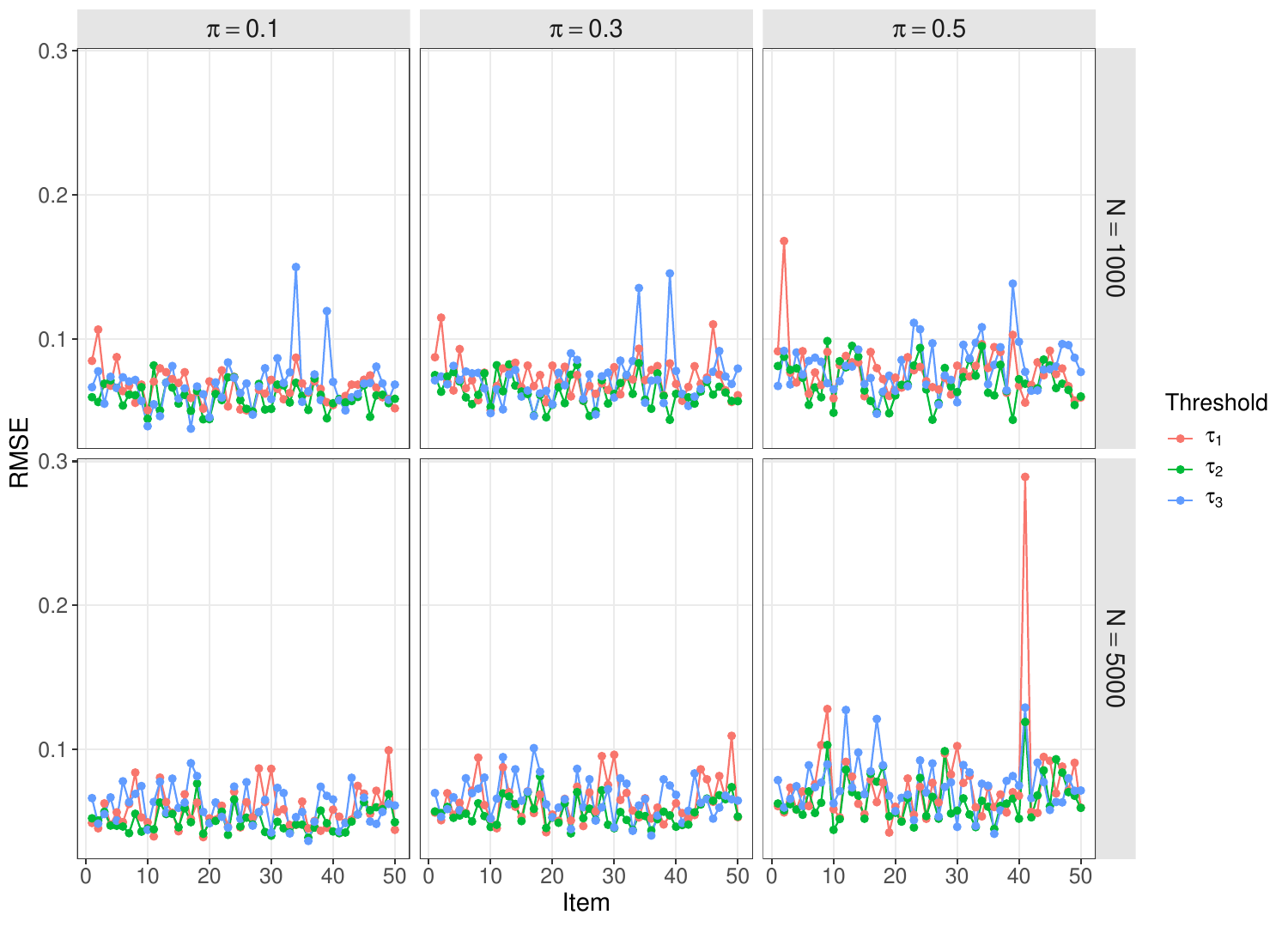}
    \caption{RMSE of threshold parameter estimates for $J=50$, $N=1000$ and $N=5000$ under the two-class case.}
\label{fig:rmse_thresholds_J50_N100050000}
\end{figure}

Figures~\ref{fig:rmse_thresholds_J15_N100050000}–\ref{fig:rmse_thresholds_J50_N100050000} summarise RMSE for the threshold parameters across \(J\), \(\pi\), and both sample sizes. For most items and thresholds, RMSE values are small, typically in the range 0.04-0.10, and show only modest variation with \(\pi\). Increasing \(J\) leads to a slight increase in the spread of RMSEs across thresholds, but there is no systematic deterioration of performance as the design becomes larger or as the class proportions become more balanced. As with the item slopes, a small number of thresholds show larger RMSEs in the most demanding settings, indicating occasional difficulties in recovering individual cutpoints, but the overall recovery of the threshold structure remains accurate.

\subsection{Respondent Classification Results}

Table~\ref{tab:respondent-two-group-merged} summarises respondent classification performance in the two-class model. 
Classification error decreases with the number of items, and AUC values increase correspondingly. 
Noticeably, for all considered class proportions, the proposed method achieves accuracy close to the oracle benchmark, with only minor loss in AUC. 
Increasing \(N\) from 1{,}000 to 5{,}000 further stabilises performance, although the gain is smaller than the improvement obtained from increasing \(J\).

Results for the three-class model are shown in Table~\ref{tab:respondent-three-groups-merged}. 
As expected, classification is more difficult in this setting, and error rates are consequently higher. 
Nevertheless, performance improves with number of items: classification error falls from roughly 0.43 at \(J=15\) to around 0.35 at \(J=50\) for \(N=1{,}000\), with a similar pattern for \(N=5{,}000\). 
Both \(\mathrm{AUC}^{(2)}\) and \(\mathrm{AUC}^{(3)}\) increase with \(J\), reaching values above 0.80 for the largest designs. 
Oracle performance is uniformly better but only modestly so, indicating that the model is able to recover latent class membership with good precision when the item set is sufficiently large.

\begin{table}[H]
\centering
\caption{Respondent classification performance for the two-group case for $N = 1{,}000$ and $N = 5{,}000$, across $J$ and $\pi$ values. The oracle performance is presented within parenthesis.}
\label{tab:respondent-two-group-merged}
\small
\begin{tabular}{llccc|ccc}
\toprule
& & \multicolumn{3}{c}{$N = 1{,}000$} & \multicolumn{3}{c}{$N = 5{,}000$} \\
\cmidrule(lr){3-5} \cmidrule(lr){6-8}
$\pi$ & Metric & $J=15$ & $J=25$ & $J=50$ & $J=15$ & $J=25$ & $J=50$ \\
\midrule

\multirow{2}{*}{0.1}
& Classification error &
\begin{tabular}{c}0.101\\(0.100)\end{tabular} &
\begin{tabular}{c}0.100\\(0.098)\end{tabular} &
\begin{tabular}{c}0.093\\(0.091)\end{tabular} &
\begin{tabular}{c}0.100\\(0.100)\end{tabular} &
\begin{tabular}{c}0.099\\(0.097)\end{tabular} &
\begin{tabular}{c}0.091\\(0.089)\end{tabular} \\
& AUC &
\begin{tabular}{c}0.775\\(0.780)\end{tabular} &
\begin{tabular}{c}0.798\\(0.814)\end{tabular} &
\begin{tabular}{c}0.865\\(0.874)\end{tabular} &
\begin{tabular}{c}0.769\\(0.776)\end{tabular} &
\begin{tabular}{c}0.820\\(0.827)\end{tabular} &
\begin{tabular}{c}0.864\\(0.874)\end{tabular} \\

\addlinespace

\multirow{2}{*}{0.3}
& Classification error &
\begin{tabular}{c}0.270\\(0.261)\end{tabular} &
\begin{tabular}{c}0.262\\(0.241)\end{tabular} &
\begin{tabular}{c}0.216\\(0.198)\end{tabular} &
\begin{tabular}{c}0.270\\(0.260)\end{tabular} &
\begin{tabular}{c}0.247\\(0.231)\end{tabular} &
\begin{tabular}{c}0.214\\(0.197)\end{tabular} \\
& AUC &
\begin{tabular}{c}0.772\\(0.781)\end{tabular} &
\begin{tabular}{c}0.793\\(0.813)\end{tabular} &
\begin{tabular}{c}0.865\\(0.875)\end{tabular} &
\begin{tabular}{c}0.766\\(0.777)\end{tabular} &
\begin{tabular}{c}0.815\\(0.828)\end{tabular} &
\begin{tabular}{c}0.860\\(0.873)\end{tabular} \\

\addlinespace

\multirow{2}{*}{0.5}
& Classification error &
\begin{tabular}{c}0.308\\(0.290)\end{tabular} &
\begin{tabular}{c}0.297\\(0.267)\end{tabular} &
\begin{tabular}{c}0.247\\(0.209)\end{tabular} &
\begin{tabular}{c}0.315\\(0.295) \end{tabular} &
\begin{tabular}{c}0.290\\(0.251)\end{tabular} &
\begin{tabular}{c}0.248\\(0.209)\end{tabular} \\
& AUC &
\begin{tabular}{c}0.773\\(0.780)\end{tabular} &
\begin{tabular}{c}0.788\\(0.807)\end{tabular} &
\begin{tabular}{c}0.845\\(0.876)\end{tabular} &
\begin{tabular}{c}0.765\\(0.777) \end{tabular} &
\begin{tabular}{c}0.807\\(0.827)\end{tabular} &
\begin{tabular}{c}0.858\\(0.875)\end{tabular} \\

\bottomrule
\end{tabular}
\end{table}

\begin{table}[H]
\centering
\caption{Respondent classification performance for the three-group case for $N = 1{,}000$ and $N = 5{,}000$, across different $J$ values. The oracle performance is presented within parenthesis.}
\label{tab:respondent-three-groups-merged}
\small
\begin{tabular}{lccc|ccc}
\toprule
& \multicolumn{3}{c}{$N = 1{,}000$} & \multicolumn{3}{c}{$N = 5{,}000$} \\
\cmidrule(lr){2-4} \cmidrule(lr){5-7}
Metric & $J=15$ & $J=25$ & $J=50$ & $J=15$ & $J=25$ & $J=50$ \\
\midrule

Classification error &
\begin{tabular}{c}0.431\\(0.403)\end{tabular} &
\begin{tabular}{c}0.399\\(0.375)\end{tabular} &
\begin{tabular}{c}0.348\\(0.329)\end{tabular} &
\begin{tabular}{c}0.425\\(0.403)\end{tabular} &
\begin{tabular}{c}0.396\\(0.375)\end{tabular} &
\begin{tabular}{c}0.356\\(0.332)\end{tabular} \\

$\text{AUC}^{(2)}$ &
\begin{tabular}{c}0.692\\(0.708)\end{tabular} &
\begin{tabular}{c}0.730\\(0.736)\end{tabular} &
\begin{tabular}{c}0.786\\(0.807)\end{tabular} &
\begin{tabular}{c}0.690\\(0.707)\end{tabular} &
\begin{tabular}{c}0.718\\(0.738)\end{tabular} &
\begin{tabular}{c}0.782\\(0.804)\end{tabular} \\

$\text{AUC}^{(3)}$ &
\begin{tabular}{c}0.769\\(0.779)\end{tabular} &
\begin{tabular}{c}0.787\\(0.811)\end{tabular} &
\begin{tabular}{c}0.810\\(0.828)\end{tabular} &
\begin{tabular}{c}0.769\\(0.778)\end{tabular} &
\begin{tabular}{c}0.802\\(0.811)\end{tabular} &
\begin{tabular}{c}0.805\\(0.825)\end{tabular} \\

\bottomrule
\end{tabular}
\end{table}


\subsection{Item Classification Results}

Table~\ref{tab:classification-item-merged} reports item classification accuracy in the two-class case. 
True positive rates for non-DIF items are essentially one across all conditions, and recovery of DIF items in the second class remains high, often above 0.95. 
False positive rates are small, though they increase slightly when \(\pi\) is large and \(J\) grows, reflecting greater overlap between groups when they are more balanced.

The three-class results in Table~\ref{tab:classification-item-three-groups-merged} show similarly strong recovery. 
True positive rates for DIF items in both additional classes are close to one for all settings, including the smaller sample size. 
False positives are somewhat higher than in the two-class case, particularly when \(J\) is large and \(N=1{,}000\), but remain moderate overall.

\begin{table}[H]
\centering
\caption{Item classification performance for the two-group case, for $N = 1{,}000$ and $N = 5{,}000$, across $J$ and $\pi$ values.
$\text{TPR}_1$ and $\text{FPR}_1$ refer to detection of uniform DIF effects ($\delta$), while
$\text{TPR}_2$ and $\text{FPR}_2$ refer to detection of non-uniform DIF effects ($\delta_2$).}

\label{tab:classification-item-merged}
\small
\begin{tabular}{llccc|ccc}
\toprule
& & \multicolumn{3}{c}{$N = 1{,}000$} & \multicolumn{3}{c}{$N = 5{,}000$} \\
\cmidrule(lr){3-5} \cmidrule(lr){6-8}
$\pi$ & Metric & $J=15$ & $J=25$ & $J=50$ & $J=15$ & $J=25$ & $J=50$ \\
\midrule

\multirow{4}{*}{0.1}
& $\text{TPR}_1$ & 1 & 1 & 1 & 1 & 1 & 1 \\
& $\text{TPR}_2$ & 0.992 & 0.960 & 0.996 & 1 & 0.996 & 1 \\
& $\text{FPR}_1$ & 0.020 & 0.021 & 0.028 & 0.024 & 0.027 & 0.023 \\
& $\text{FPR}_2$ & 0.008 & 0.011 & 0.015 & 0.012 & 0.019 & 0.025 \\

\addlinespace

\multirow{4}{*}{0.3}
& $\text{TPR}_1$ & 1 & 1 & 1 & 1 & 1 & 1 \\
& $\text{TPR}_2$ & 1 & 1 & 1 & 1 & 1 & 0.998 \\
& $\text{FPR}_1$ & 0.048 & 0.072 & 0.048 & 0.056 & 0.096 & 0.060 \\
& $\text{FPR}_2$ & 0.068 & 0.083 & 0.043 & 0.076 & 0.077 & 0.105 \\

\addlinespace

\multirow{4}{*}{0.5}
& $\text{TPR}_1$ & 1 & 1 & 1 & 1 & 1 & 1 \\
& $\text{TPR}_2$ & 1 & 1 & 1 & 1 & 0.996 & 1 \\
& $\text{FPR}_1$ & 0.036 & 0.056 & 0.097 & 0.124 & 0.099 & 0.092 \\
& $\text{FPR}_2$ & 0.044 & 0.091 & 0.137 & 0.132 & 0.136 & 0.181 \\
\bottomrule
\end{tabular}
\end{table}

\begin{table}[H]
\centering
\caption{Item classification performance metrics for the three-group case for $N = 1{,}000$ and $N = 5{,}000$, across different $J$ values.
$\text{TPR}^{(g)}_1$ and $\text{FPR}^{(g)}_1$ denote the true and false positive rates for detecting uniform DIF effects ($\delta$),
and $\text{TPR}^{(g)}_2$ and $\text{FPR}^{(g)}_2$ denote the corresponding rates for non-uniform DIF effects ($\delta_2$),
with superscript $(g)$ indicating the comparison group ($g=2,3$) relative to the reference group.}
\label{tab:classification-item-three-groups-merged}
\small
\begin{tabular}{lccc|ccc}
\toprule
& \multicolumn{3}{c}{$N = 1{,}000$} & \multicolumn{3}{c}{$N = 5{,}000$} \\
\cmidrule(lr){2-4} \cmidrule(lr){5-7}
Metric & $J=15$ & $J=25$ & $J=50$ & $J=15$ & $J=25$ & $J=50$ \\
\midrule

$\text{TPR}^{(2)}_1$ & 1 & 1 & 1 & 1 & 1 & 1 \\
$\text{TPR}^{(2)}_2$ & 1 & 1 & 1 & 1 & 1 & 1 \\
$\text{FPR}^{(2)}_1$ & 0.036 & 0.061 & 0.073 & 0.084 & 0.083 & 0.123 \\
$\text{FPR}^{(2)}_2$ & 0.052 & 0.072 & 0.075 & 0.084 & 0.112 & 0.147 \\

\addlinespace\hdashline\addlinespace

$\text{TPR}^{(3)}_1$ & 1 & 1 & 1 & 1 & 1 & 1 \\
$\text{TPR}^{(3)}_2$ & 1 & 1 & 1 & 1 & 1 & 1 \\
$\text{FPR}^{(3)}_1$ & 0.044 & 0.069 & 0.063 & 0.076 & 0.085 & 0.151 \\
$\text{FPR}^{(3)}_2$ & 0.076 & 0.051 & 0.080 & 0.100 & 0.083 & 0.177 \\
\bottomrule
\end{tabular}
\end{table}

\subsection{Structural Parameters}

Structural parameter estimation for the two-class case is presented in Table~\ref{tab:global-structural-combined-merged}. 
Bias in class proportions \(\pi\) is small in all settings, and RMSE decreases as \(J\) increases or as \(N\) grows. 
Estimates of \(\mu\) and $\sigma$ are also close to unbiased, though uncertainty is larger when the classes are of similar size \((\pi = 0.5)\).

Table~\ref{tab:global-structural-three-groups-merged} shows results for the three-class model. Bias in the mixing proportions are small, and RMSE again declines with increasing number of items. 
Estimates of \(\mu_2\) and \(\mu_3\) show small negative bias for \(N = 1{,}000\), with clear improvement for \(N = 5{,}000\). 
Variance parameters \(\sigma_2\) and \(\sigma_3\) show similar behaviour, with larger dispersion at small \(J\) and smaller RMSE for larger designs.

\begin{table}[H]
\centering
\caption{Mean bias and RMSE (within parenthesis) for structural model parameters for $N = 1{,}000$ and $N = 5{,}000$ across $J$ and $\pi$ values in the two-group case.}
\label{tab:global-structural-combined-merged}
\small
\begin{tabular}{llccc|ccc}
\toprule
& & \multicolumn{3}{c}{$N = 1{,}000$} & \multicolumn{3}{c}{$N = 5{,}000$} \\
\cmidrule(lr){3-5} \cmidrule(lr){6-8}
$\pi$ & Parameter & $J=15$ & $J=25$ & $J=50$ & $J=15$ & $J=25$ & $J=50$ \\
\midrule

\multirow{3}{*}{0.1}
& $\pi$    &
\begin{tabular}{c}-0.009\\(0.045)\end{tabular} &
\begin{tabular}{c}-0.008\\(0.037)\end{tabular} &
\begin{tabular}{c}-0.007\\(0.022)\end{tabular} &
\begin{tabular}{c}-0.011\\(0.025)\end{tabular} &
\begin{tabular}{c}-0.012\\(0.028)\end{tabular} &
\begin{tabular}{c}-0.010\\(0.018)\end{tabular} \\
& $\mu$    &
\begin{tabular}{c}-0.027\\(0.121)\end{tabular} &
\begin{tabular}{c}0.028\\(0.106)\end{tabular} &
\begin{tabular}{c}-0.025\\(0.104)\end{tabular} &
\begin{tabular}{c}0.007\\(0.092)\end{tabular} &
\begin{tabular}{c}0.016\\(0.096)\end{tabular} &
\begin{tabular}{c}0.010\\(0.110)\end{tabular} \\
& $\sigma$ &
\begin{tabular}{c}0.005\\(0.086)\end{tabular} &
\begin{tabular}{c}-0.012\\(0.114)\end{tabular} &
\begin{tabular}{c}-0.007\\(0.109)\end{tabular} &
\begin{tabular}{c}0.015\\(0.106)\end{tabular} &
\begin{tabular}{c}-0.021\\(0.099)\end{tabular} &
\begin{tabular}{c}0.028\\(0.112)\end{tabular} \\

\addlinespace

\multirow{3}{*}{0.3}
& $\pi$    &
\begin{tabular}{c}-0.011\\(0.050)\end{tabular} &
\begin{tabular}{c}-0.043\\(0.066)\end{tabular} &
\begin{tabular}{c}-0.023\\(0.050)\end{tabular} &
\begin{tabular}{c}-0.025\\(0.039)\end{tabular} &
\begin{tabular}{c}-0.035\\(0.050)\end{tabular} &
\begin{tabular}{c}-0.036\\(0.049)\end{tabular} \\
& $\mu$    &
\begin{tabular}{c}-0.053\\(0.164)\end{tabular} &
\begin{tabular}{c}0.044\\(0.185)\end{tabular} &
\begin{tabular}{c}-0.020\\(0.146)\end{tabular} &
\begin{tabular}{c}0.005\\(0.120)\end{tabular} &
\begin{tabular}{c}0.027\\(0.131)\end{tabular} &
\begin{tabular}{c}0.020\\(0.147)\end{tabular} \\
& $\sigma$ &
\begin{tabular}{c}-0.012\\(0.118)\end{tabular} &
\begin{tabular}{c}-0.038\\(0.166)\end{tabular} &
\begin{tabular}{c}-0.032\\(0.157)\end{tabular} &
\begin{tabular}{c}0.007\\(0.130)\end{tabular} &
\begin{tabular}{c}-0.043\\(0.142)\end{tabular} &
\begin{tabular}{c}0.014\\(0.151)\end{tabular} \\

\addlinespace

\multirow{3}{*}{0.5}
& $\pi$    &
\begin{tabular}{c}-0.028\\(0.074)\end{tabular} &
\begin{tabular}{c}-0.062\\(0.089)\end{tabular} &
\begin{tabular}{c}-0.064\\(0.105)\end{tabular} &
\begin{tabular}{c} -0.049 \\(0.065) \end{tabular} &
\begin{tabular}{c}-0.076\\(0.102)\end{tabular} &
\begin{tabular}{c}-0.076\\(0.094)\end{tabular} \\
& $\mu$    &
\begin{tabular}{c}-0.023\\(0.188)\end{tabular} &
\begin{tabular}{c}0.045\\(0.176)\end{tabular} &
\begin{tabular}{c}-0.061\\(0.320)\end{tabular} &
\begin{tabular}{c} 0.011 \\(0.138) \end{tabular} &
\begin{tabular}{c}0.016\\(0.171)\end{tabular} &
\begin{tabular}{c}0.007\\(0.182)\end{tabular} \\
& $\sigma$ &
\begin{tabular}{c}-0.005\\(0.130)\end{tabular} &
\begin{tabular}{c}-0.038\\(0.187)\end{tabular} &
\begin{tabular}{c}-0.043\\(0.169)\end{tabular} &
\begin{tabular}{c} 0.011 \\(0.142) \end{tabular} &
\begin{tabular}{c}-0.071\\(0.189)\end{tabular} &
\begin{tabular}{c}0.015\\(0.201)\end{tabular} \\
\bottomrule
\end{tabular}
\end{table}

\begin{table}[H]
\centering
\caption{Mean bias and RMSE (within parenthesis) in parenthesis for structural model parameters in the three-group case for $N=1{,}000$ and $N=5{,}000$.}
\label{tab:global-structural-three-groups-merged}
\small
\begin{tabular}{lccc|ccc}
\toprule
& \multicolumn{3}{c}{$N = 1{,}000$} & \multicolumn{3}{c}{$N = 5{,}000$} \\
\cmidrule(lr){2-4} \cmidrule(lr){5-7}
Parameter & $J=15$ & $J=25$ & $J=50$ & $J=15$ & $J=25$ & $J=50$\\
\midrule
$\pi_1$ &
\begin{tabular}{@{}c@{}}-0.014\\(0.067)\end{tabular} &
\begin{tabular}{@{}c@{}}-0.009\\(0.052)\end{tabular} &
\begin{tabular}{@{}c@{}}-0.003\\(0.037)\end{tabular} &
\begin{tabular}{@{}c@{}}-0.006\\(0.047)\end{tabular} &
\begin{tabular}{@{}c@{}}0.006\\(0.043)\end{tabular} &
\begin{tabular}{@{}c@{}}0.009\\(0.042)\end{tabular} \\
$\pi_2$ &
\begin{tabular}{@{}c@{}}0.037\\(0.120)\end{tabular} &
\begin{tabular}{@{}c@{}}0.042\\(0.078)\end{tabular} &
\begin{tabular}{@{}c@{}}0.030\\(0.051)\end{tabular} &
\begin{tabular}{@{}c@{}}0.026\\(0.098)\end{tabular} &
\begin{tabular}{@{}c@{}}-0.001\\(0.068)\end{tabular} &
\begin{tabular}{@{}c@{}}0.002\\(0.046)\end{tabular} \\
$\pi_3$ &
\begin{tabular}{@{}c@{}}-0.023\\(0.085)\end{tabular} &
\begin{tabular}{@{}c@{}}-0.033\\(0.056)\end{tabular} &
\begin{tabular}{@{}c@{}}-0.027\\(0.055)\end{tabular} &
\begin{tabular}{@{}c@{}}-0.019\\(0.083)\end{tabular} &
\begin{tabular}{@{}c@{}}-0.005\\(0.059)\end{tabular} &
\begin{tabular}{@{}c@{}}-0.011\\(0.044)\end{tabular} \\
\hdashline
$\mu_2$ &
\begin{tabular}{@{}c@{}}-0.036\\(0.139)\end{tabular} &
\begin{tabular}{@{}c@{}}0.014\\(0.118)\end{tabular} &
\begin{tabular}{@{}c@{}}0.034\\(0.142)\end{tabular} &
\begin{tabular}{@{}c@{}}-0.025\\(0.141)\end{tabular} &
\begin{tabular}{@{}c@{}}-0.010\\(0.116)\end{tabular} &
\begin{tabular}{@{}c@{}}-0.001\\(0.132)\end{tabular} \\
$\mu_3$ &
\begin{tabular}{@{}c@{}}-0.028\\(0.111)\end{tabular} &
\begin{tabular}{@{}c@{}}-0.022\\(0.108)\end{tabular} &
\begin{tabular}{@{}c@{}}-0.011\\(0.099)\end{tabular} &
\begin{tabular}{@{}c@{}}0.000\\(0.111)\end{tabular} &
\begin{tabular}{@{}c@{}}0.048\\(0.169)\end{tabular} &
\begin{tabular}{@{}c@{}}-0.026\\(0.090)\end{tabular} \\
\hdashline
$\sigma_2$ &
\begin{tabular}{@{}c@{}}0.014\\(0.103)\end{tabular} &
\begin{tabular}{@{}c@{}}-0.045\\(0.126)\end{tabular} &
\begin{tabular}{@{}c@{}}-0.059\\(0.146)\end{tabular} &
\begin{tabular}{@{}c@{}}0.011\\(0.132)\end{tabular} &
\begin{tabular}{@{}c@{}}-0.005\\(0.124)\end{tabular} &
\begin{tabular}{@{}c@{}}-0.018\\(0.155)\end{tabular} \\
$\sigma_3$ &
\begin{tabular}{@{}c@{}}-0.015\\(0.099)\end{tabular} &
\begin{tabular}{@{}c@{}}-0.056\\(0.113)\end{tabular} &
\begin{tabular}{@{}c@{}}0.013\\(0.110)\end{tabular} &
\begin{tabular}{@{}c@{}}-0.019\\(0.111)\end{tabular} &
\begin{tabular}{@{}c@{}}-0.019\\(0.093)\end{tabular} &
\begin{tabular}{@{}c@{}}0.003\\(0.135)\end{tabular} \\
\bottomrule
\end{tabular}
\end{table}

{A small number of RMSE values for \(\mu_k\) and \(\sigma_k\) do not decrease when moving from \(N=1{,}000\) to \(N=5{,}000\). We attribute this to the finite-sample behaviour of the proposed mixture model rather than to a systematic loss of accuracy. 
Unlike the class proportions, the class specific means and standard deviations are estimated indirectly through the ordinal  response model and jointly with item parameters and DIF effects. Their estimation is therefore sensitive to class overlap, DIF structure, and convergence to local optima.  In addition, the reported RMSEs are Monte Carlo summaries and may show small non-monotonic fluctuations across simulation conditions.}

{As an additional robustness check, Appendix~\ref{app:class-number-sensitivity} reports a sensitivity analysis examining the effect of misspecifying the number of latent classes. The results show that underfitting the number of classes prevents recovery of the latent DIF structure, whereas moderate overfitting gives similar classification performance but increases false positive DIF selections and is disfavoured by BIC.}


\section{Discussion}\label{sec:Discussion}

This paper developed a regularised hybrid latent-class item response framework for detecting measurement non-invariance in ordinal-scale instruments without requiring predefined comparison groups or anchor items. By combining a proportional-odds IRT formulation with latent mixture modelling and an $\ell_1$ penalty on DIF effects, the framework enables simultaneous discovery of latent subgroups and identification of both uniform and non-uniform DIF at the item level. This feature distinguishes the method from traditional DIF approaches, which require anchor specification, and from most penalised DIF methods that rely on known grouping structures, uniform DIF only and/or binary data. The proposed estimation strategy is computationally feasible through a modified EM algorithm with proximal updates, and the confirmatory post-selection refitting step recovers unbiased effect estimates for interpretability.

The empirical illustration shows that the proposed model can reveal substantively interpretable patterns of measurement heterogeneity that would be obscured under a single-population specification. The two-class solution identifies a relatively small subgroup of respondents characterised by systematically lower endorsement levels and reduced latent variability, particularly on items related to competitiveness, directness, and socially situated expressions of analytical thinking. Importantly, this heterogeneity is expressed not only through overall shifts in endorsement but also through both uniform and non-uniform DIF. The latent classes are not well explained by standard socio-demographic variables. Age, education, and most background characteristics show little to no association with class membership, suggesting that the detected heterogeneity does not simply reflect known external groupings. Gender, native language, and handedness exhibit statistically detectable associations, providing contextual cues of the latent structure. Notably, when mixture structure is ignored, gender differences are absorbed into mean shifts on the latent scale, whereas the mixture model reallocates this variation into differences in class membership probabilities. These findings are thus consistent with the proposed model. 

{At the same time, the latent classes and DIF effects should be interpreted as model-based quantities rather than as direct evidence of substantive item bias. Because class membership is not externally observed but inferred from the same response data used to estimate the item parameters, the mixture components need not correspond to substantively distinct populations. They may also reflect unmodelled heterogeneity in the latent trait distribution, response styles, local dependence, multidimensionality, or other forms of behavioural variation. Similarly, non-zero estimates of \(\delta_{1jk}\) and \(\delta_{2jk}\) indicate class-specific departures from measurement invariance under the fitted model, but should not by themselves be interpreted as definitive evidence of item bias. {In particular, the slope DIF parameters \(\delta_{2jk}\) should be interpreted as class-specific slope differences under the fitted latent mixture model, rather than as traditional non-uniform DIF with respect to externally observed groups}. Substantive interpretation should therefore be supported by additional evidence, such as item content review, external covariates, model-fit diagnostics, and sensitivity analyses using alternative numbers of classes or penalty values.}

{A related interpretational issue concerns the role of the sparsity assumption in anchoring the measurement scale. 
In the absence of externally specified anchor items, the class-specific latent distribution parameters and the class-specific DIF parameters are only separable under additional restrictions. The proposed regularised estimator resolves this ambiguity by favouring solutions in which most class-specific DIF effects are zero, thereby allowing the invariant items selected by the penalty to define the common measurement scale. This provides a data-driven alternative to pre-specifying anchor items, but it also means that interpretation depends on the adequacy of the sparsity assumption. If DIF is widespread rather than sparse, the model may attribute part of the non-invariance to differences in the latent class distribution, or may favour additional mixture components to accommodate residual item-level heterogeneity. In such cases, latent classes may partly absorb measurement non-invariance, and the selected number of classes may overstate substantive population heterogeneity.  For this reason, the estimated class structure and the detected DIF effects should be interpreted jointly.}

Simulation studies further validated the proposed framework across a range of realistic experimental conditions. For two latent classes, item discriminations and threshold parameters were estimated with low RMSE, and DIF parameters exhibited a strong recovery pattern: near-zero error for invariant items and increased RMSE only for those generated with DIF. Respondent classification was highly accurate, approaching oracle performance. The method showed strong DIF detection performance with low false-positive rates. The three-class experiments revealed the expected increase in difficulty, yet classification accuracy, structural parameter recovery, and item allocation were all robust, improving steadily with number of items and sample size.

Several limitations should be considered for future work. {First, the framework represents heterogeneity using a finite number of latent classes. This provides a flexible discrete approximation to population heterogeneity, but if the underlying heterogeneity is continuous, the estimated classes may be better understood as a parsimonious summary of response variation rather than as distinct subgroups. Appendix~\ref{app:class-number-sensitivity} provides a sensitivity analysis examining the effect of misspecifying the number of latent classes. In the setting considered, BIC favoured the correctly specified two-class model; underfitting prevented recovery of the latent DIF structure, whereas moderate overfitting gave similar classification performance but increased false positive DIF selections.} Second, the present implementation assumes a unidimensional latent trait. Extensions to multidimensional structures would allow modelling of more complex psychological constructs but require careful identifiability analysis and more demanding computation. Third, the proportional-odds structure assumes threshold shifts are class-invariant, while DIF enters only through location and slope modifications. Allowing class-specific thresholds may capture a broader range of response behaviour but would substantially increase parameter dimensionality and inference complexity. {If the proportional-odds assumption is violated, item-level misfit may be partly absorbed by the latent classes or by the penalised DIF effects}. {Fourth, this paper focuses on estimation and model selection rather than inference. Although the confirmatory refit removes shrinkage bias, it does not provide valid post-selection standard errors because it conditions on the active DIF structure selected by the penalised estimator. Developing selective-inference, bootstrap, or debiased-estimation procedures for this setting is therefore an important direction for future work.} Finally, integrating auxiliary covariates directly into the latent-class membership model may improve interpretability and would be consistent with explanatory analysis, rather than post-hoc association testing.

Despite these challenges, the results indicate that the proposed approach provides a flexible framework for assessing measurement invariance in ordinal survey instruments when comparison groups are unknown. The method scales to realistic item sets, recovers latent structure with good precision, and identifies non-invariant items with high accuracy. We anticipate that this framework will be useful in domains where population heterogeneity is expected but difficult to formalise a priori.

\bibliography{bibliography}

\newpage

\section{Supplementary Material}\label{sec:Appendix}

\subsection{Items analysed from the FTI}

\begin{enumerate}
    \item I understand complex machines easily.
    \item I enjoy competitive conversations.
    \item I am intrigued by rules and patterns that govern systems.
    \item I am more analytical and logical than most people.
    \item I pursue intellectual topics thoroughly and regularly.
    \item I am able to solve problems without letting emotion get in the way.
    \item I like to figure out how things work.
    \item I am tough-minded.
    \item Debating is a good way to match my wits with others.
    \item I have no trouble making a choice, even when several alternatives seem equally good at first.
    \item When I buy a new machine (like a camera, computer or car), I want to know all of its technical features.
    \item I like to avoid the nuances and say exactly what I mean.
    \item I think it is important to be direct.
    \item When making a decision, I like to stick to the facts rather than letting emotions influence the decisions.
\end{enumerate}

\subsection{Forced-Choice Items}
\begin{enumerate}
\item 1 = I constantly seek new adventures. 2 = I generally prefer to do familiar things.
\item 1 = I'm interested in all kinds of different people. 2 = I am interested in people who share my deepest interests.
\item 1 = I'm not very introspective; I like to look out not in. 2 = I'm very introspective; I'm interested in deeply understanding others.
\item 1 = I tend to be cautious in my work and thinking. 2 = I tend to be daring in my work and thinking.
\item 1 = I tend to think concretely; I only trust the facts. 2 = I tend to be imaginative and listen to my intuition.
\item 1 = I tend to be tough minded. 2 = I tend to be tender hearted.
\end{enumerate}

\newpage

\subsection{Simulation results for the three-group case}

\begin{figure}[H]
    \centering
    \includegraphics[width=1\linewidth]{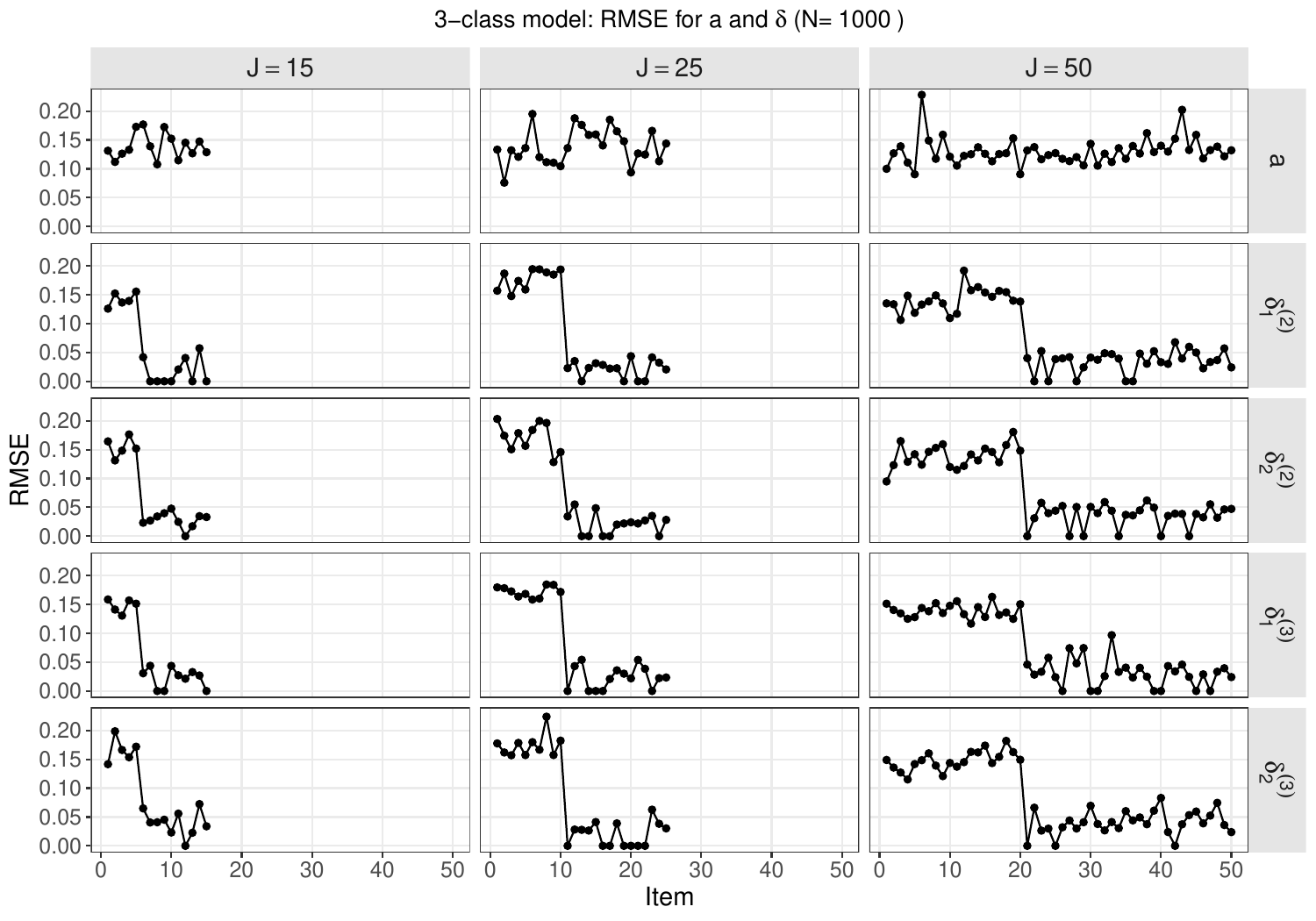}
    \caption{RMSE of the item parameter estimates for $N=1,000$ and $J=\{15, 25, 50\}$ for the three-class case.}
    \label{fig:RMSE_threeclass_item_params}
\end{figure}

\begin{figure}
    \centering
    \includegraphics[width=1\linewidth]{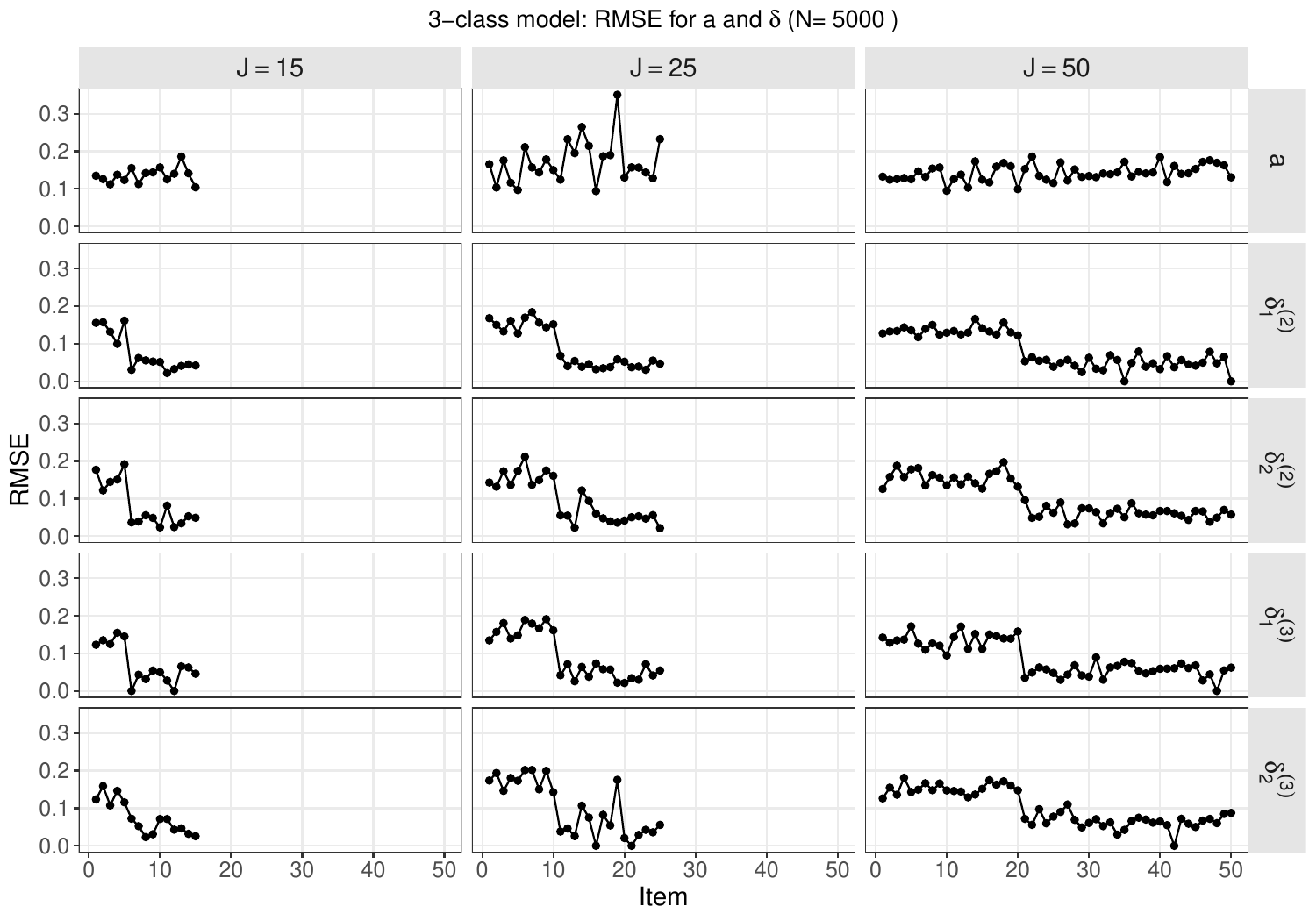}
    \caption{RMSE of the threshold parameter estimates for $N=1,000$ and $J=\{15, 25, 50\}$ for the three-class case.}
    \label{fig:RMSE_threeclass_thresholds}
\end{figure}

\begin{figure}
    \centering
    \includegraphics[width=1\linewidth]{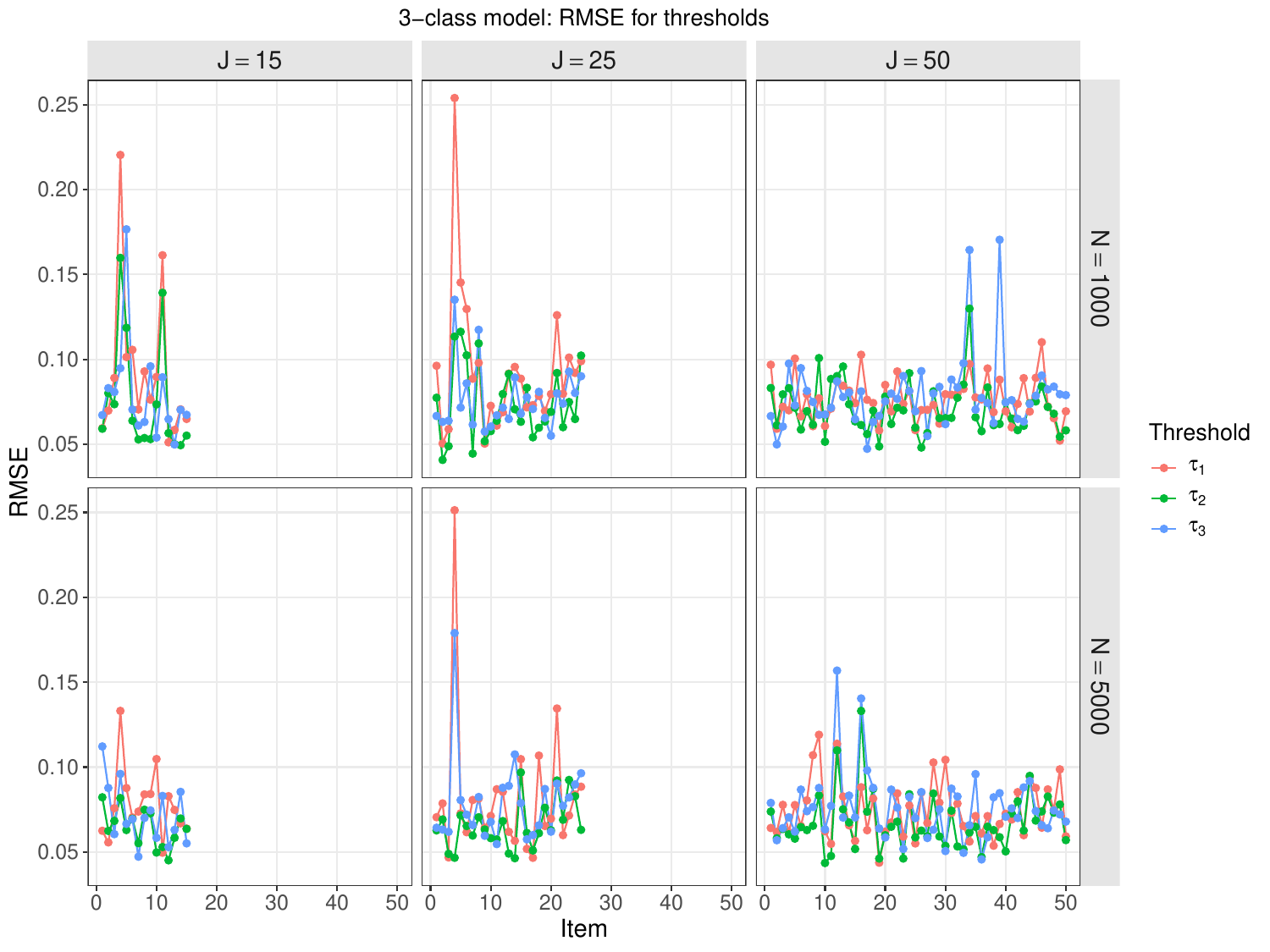}
    \caption{RMSE for the threshold parameters in the three-group case.}
    \label{fig:threeclass_threshold_rmse}
\end{figure}

\newpage

\section{Sensitivity analysis for misspecifying the number of latent classes}
\label{app:class-number-sensitivity}

A concern with latent-class DIF models is that the estimated DIF structure may depend on the assumed number of latent classes. If too few classes are fitted, genuine heterogeneity may be absorbed into item parameters or left unexplained. Conversely, if too many classes are fitted, the model may split an existing latent group into multiple components, potentially increasing estimation variability and complicating interpretation. To assess the sensitivity of the proposed procedure to this issue, we conducted an additional simulation study in which the data-generating model contained two latent classes, while the fitted models assumed one, two, or three latent classes, respectively.

The data were generated from the ordinal latent-class DIF model with two latent classes. We used \(N=1000\) respondents, \(J=15\) items, and four ordered response categories per item. Five items were generated to display both uniform and non-uniform DIF. The latent trait distribution in the reference class was standard normal, while the non-reference class had mean \(\mu=1.0\) and standard deviation \(\sigma=0.8\). The class proportion was set to \(\pi=0.5\). For each simulated data set, we fitted models with one, two, and three latent classes. For each fitted number of classes, the tuning parameter was selected by BIC over a grid of \(\lambda\) values, after which the selected active DIF structure was refitted without penalisation, as described in Section~\ref{sec:statistical-framework}. The sensitivity analysis used 100 replications.

Table~\ref{tab:class-number-sensitivity} summarises the results. As expected, fitting a one-class model fails to recover the latent class structure: the classification error is close to chance, the AUC is 0.50, and no DIF effects are identified. This reflects the fact that a one-class model cannot represent class-specific measurement non-invariance. The correctly specified two-class model gives the lowest mean BIC and recovers both types of DIF well. In particular, the true positive rates for both uniform and non-uniform DIF are equal to 1.00, while the false positive rates remain low. The three-class model gives classification performance similar to the two-class model, but has a higher mean BIC and slightly higher false positive rates. This suggests that overfitting the number of classes does not substantially improve recovery of the true latent structure, but may introduce additional spurious DIF selections and reduce interpretability.

\begin{table}[htbp]
\centering
\caption{Sensitivity of the proposed method to misspecification of the number of latent classes. The true data-generating model contains two latent classes. Entries are means across 100 replications.}
\label{tab:class-number-sensitivity}
\resizebox{\textwidth}{!}{
\begin{tabular}{lccccccc}
\hline
Fitted classes & BIC & Class error & AUC & TPR \(\delta_1\) & FPR \(\delta_1\) & TPR \(\delta_2\) & FPR \(\delta_2\) \\
\hline
1 & 33672.6 & 0.491 & 0.500 & 0.000 & 0.000 & 0.000 & 0.000 \\
2 & 33599.5 & 0.314 & 0.766 & 1.000 & 0.040 & 1.000 & 0.060 \\
3 & 33623.8 & 0.310 & 0.764 & 1.000 & 0.080 & 1.000 & 0.120 \\
\hline
\end{tabular}
}
\end{table}

\end{document}